%% file: weinmann.tex
\documentclass[letterpaper]{mn2e}
\input{psfig}
\input{macros_wl}

\def\lesssim{\mathrel{\hbox{\rlap{\hbox{\lower4pt\hbox{$\sim$}}}\hbox{$<$}}}}
\def\gtrsim{\mathrel{\hbox{\rlap{\hbox{\lower4pt\hbox{$\sim$}}}\hbox{$>$}}}}

\def\aj{AJ}
\def\apj{ApJ}
\def\apjl{ApJ}
\def\apjs{ApJS}
\def\aap{A\&A}
\def\aaps{A\&AS}
\def\mnras{MNRAS}
\def\nat{Nature}
\newcommand{\fof}{{\scshape fof~}}
\newcommand{\subfind}{{\scshape subfind~}}

\usepackage{amsmath}
\usepackage{graphicx}  
\usepackage{color}

\newcommand{\comment}[1]{}

\begin{document}


\title[Abundances and red fractions of dwarf galaxies in clusters]{Dwarf galaxy populations in present-day galaxy clusters: I. Abundances and
red fractions}
\author[S. M. Weinmann, T.Lisker, Q.Guo et al.]
       {\parbox[t]{\textwidth}{
        Simone
        M. Weinmann$^{1}$\thanks{E-mail:weinmann@strw.leidenuniv.nl},
        Thorsten Lisker$^{2}$, Qi Guo$^{3,4}$, Hagen T. Meyer$^{2}$,
        Joachim Janz$^{2,5}$\thanks{Fellow of the Gottlieb Daimler and Karl Benz Foundation}
        }\\
\vspace*{3pt}\\
$^1$Leiden Observatory, Leiden University, P.O. Box 9513, 2300 RA Leiden, The Netherlands\\
$^2$Astronomisches Rechen-Institut, Zentrum f\"ur Astronomie der
  Universit\"at Heidelberg, M\"onchhofstra\ss e 12-14, 69120
  Heidelberg, Germany\\
$^3$National Astronomical Observatories, Chinese Academy of Sciences,
Beijing 100012, China\\
$^4$Institute for Computational Cosmology, Department of Physics,
University of Durham, South Road, Durham, DH1 3LE, UK\\
$^5$Division of Astronomy, Department of Physical Sciences,
  University of Oulu, P.O.\ Box 3000, FIN-90014 Oulu, Finland\\
}


\date{}

\pubyear{2011}

\maketitle

\label{firstpage}


\begin{abstract}
We  compare the  galaxy  population  in the  Virgo,  Fornax, Coma  and
Perseus cluster  to a state-of-the-art semi-analytic model, focusing  on the regime
of dwarf galaxies with luminosities from  approximately $10^8$ L$_{\odot}$ to
$10^9$ L$_{\odot}$.  
We find that the number density profiles of dwarfs
in observed clusters are reproduced reasonably well, and that the red
fractions of model clusters provide a good match to Coma and Perseus.
On the other hand, the red fraction among dwarf galaxies in 
Virgo is clearly lower than in model clusters.
We argue that this is mainly caused by the treatment of environmental
effects in the model. This explanation is supported by our finding
that the colours of central (``field'') dwarf galaxies are reproduced well, 
in contrast to previous claims.
Finally, we find that the dwarf-to-giant ratio in model clusters is too high.
This may indicate that the current model prescription for 
tidal disruption of faint galaxies is still not efficient
enough.
\end{abstract}




\section{Introduction}
\label{sec:intro}
Dwarf galaxies  in nearby  clusters display an 
intriguing variety  in their
properties  such as  colour   and  morphology,   often   accompanied  by
differences  in their kinematic  and spatial  distribution. Several
types have been distinguished, such as 
 dwarf irregulars (dIrr),  dwarf ellipticals (dE) and blue compact
dwarf galaxies (BCD) (e.g. Sandage \& Binggeli 1984).
Recent studies
have established  the presence of  even finer subclasses: dEs with
and  without  nuclei,  as well as with  and  without disk components,
were found to  have
different shapes, stellar populations, spatial and kinematic distributions (Lisker et
al.  2006, 2007,  2008, 2009; Toloba et al. 2009; Paudel et al. 2010). 
The  origin of  those  subclasses, and  the
evolution  of cluster  dwarf  galaxies  in general,  is  not yet  well
understood theoretically.  The reason for  this is partially  that the
large dynamic range between a  dwarf galaxy and its host cluster makes
it difficult to study them with    hydrodynamical    simulations.
Semi-analytic  models (SAMs,  e.g. Kauffmann  et al.  1993;  Cole et
al.  2000)  applied to  N-body  simulations  are therefore a  viable
alternative.  While SAMs have contributed significantly to our understanding of
higher mass galaxies,  they have not been exploited  much in the study
of  dwarf galaxies, with the exception of the dwarf galaxies in the Local 
Group (e.g. 
Benson et al. 2002; Macci\`{o}  et  al. 2010; Li et al. 
2010; Font et al. 2011). Some basic comparisons between dwarf galaxies
in clusters in SAMs and observations have been made by Kauffmann et al.  (1993), Springel et al. (2001) and
Benson et al. (2003), all finding that the shape of the $B$-band cluster
luminosity function is reproduced well down to faint magnitude (with the latter
two neglecting its normalization). Tully et al. (2002) have used a SAM to 
infer formation redshifts of cluster dwarf galaxies, while
Nagashima \& Yoshii (2004) and
Janz \& Lisker (2008, 2009) focus
on scaling relations in dwarf galaxies. In all these studies, SAMs have either been based on analytical
merger trees, or were limited to small volumes.\\  
The new high-resolution Millennium-II  simulation (hereafter MS-II,
Boylan-Kolchin  et  al.  2009)   now  allows  SAMs  based  on  N-body
simulations of a  cosmological volume to probe the  regime of low mass
galaxies.    The  first   such  SAM   has  recently   been  publicly
released\footnote{SQL databases containing the full galaxy data at all
  redshifts and for both  the Millennium and MS-II simulations
  are                publicly                released               at
  \texttt{http://www.mpa-garching.mpg.de/millennium}}(Guo et al. 2011,
G11 hereafter).  It has been tuned to reproduce the $z$=0 stellar mass
function  (SMF hereafter) down  to  log($M_{\rm   star}/M_{\odot})  \sim  7.5$,  andalso matches
the  $z$=0
luminosity  function (LF hereafter)
down  to $M_r  = -15$ reasonably well, making it a
promising  tool to understand  the origin  and evolution  of the
  different dwarf galaxy types.

Before more detailed studies are conducted, it  is however important
to check whether  the global properties  of cluster  dwarf galaxies,
like their  kinematic and spatial distribution,  their abundance and
their  luminosities,  agree  between the  model and  observations.  In
this paper, we therefore concentrate  on the global statistics and the
colours of dwarf  galaxies. We restrict ourselves to  the  nearby 
Virgo, Coma,
Fornax and Perseus cluster  for two reasons. First, it is only
in such nearby clusters that we can be confident of obtaining complete
samples of faint cluster galaxies  (or at least samples with very well
known  incompleteness).
   Second,   those  clusters  have  been  studied
extensively, and are often used as  benchmarks to draw conclusions about
dwarf galaxy
formation  and evolution  in general.   It is  thus important  to know
whether  or not  it  is safe  to  directly compare  those clusters  to
similar  mass  clusters  in the  SAM,   whether  some of  them  are
atypical, and whether the model fails to reproduce some of their
properties in general.

In  fact,  it  has  been  shown  that several previous  SAMs
(e.g. Croton et  al. 2006; Bower et al. 2006; Kang et al. 2005) 
overpredicted  the satellite content of
groups  and  clusters  (Weinmann  et  al.  2006b;  Liu  et
al. 2010). It has been argued that this problem is due to 
the lack of (both partial and full) tidal disruption of satellites 
in SAMs. This conclusion is supported by additional evidence 
from the presence of intracluster 
light (e.g. Zibetti et al. 2004), the need to 
reconcile the halo occupation statistics with halo merger rates
(e.g. Conroy, Ho \& White 2007; Yang, Mo \& van den Bosch 2009)
and the high metallicity
of observed satellite galaxies (Pasquali et al. 2010).
 Satellite disruption has therefore
 been included in several recent
SAMs in various ways (e.g. Benson et al. 2002; Kang \& van den Bosch 2008; Kim et al. 2009; 
Henriquez \& Thomas 2010). In the G11 SAM, 
only full disruption is included, and occurs when 
the average baryon density in the galaxy falls below that of the surrounding
dark matter (see sec. \ref{sec:env_SAM}). Possibly as a result of
this,
the number density profile of relatively massive galaxies
in clusters in G11 has been shown to agree well with
observations.
In this work, we focus on faint galaxies in massive clusters, 
for which no comparison has been made yet. 
We  find that the match  is reasonable, but
that the  SAM overpredicts the  ratio between faint  and bright
cluster galaxies, which indicates that tidal
disruption of low mass galaxies may still be underestimated.

The  colours of satellite  galaxies, on  the other  hand, can  give us
insights  into  how environment affect  star formation  in
groups  and  clusters.   For   satellites  more  massive  than  $\sim$
log$(M_{\rm   star}/M_{\odot})$=9.5,  there   are strong indications that
  the  main  mechanism  that  is  responsible  for  the
quenching  of  star  formation  is ``starvation''(Larson,  Tinsley  \&
Caldwell  1980), which  is the  gradual removal  of the  hot  gas halo
around galaxies (e.g. Weinmann et al. 2006a, 2009; Font et al. 2008; 
van den Bosch et al. 2008; van der
Wel et al. 2010; von der Linden et al. 2010).  The previous generation
of SAMs mimicked  this process by simply removing  the entire extended
gas reservoir  from a galaxy when  it became a satellite.  This led to
too high red fractions  in satellite galaxies compared to observations
(Weinmann et al. 2006b; Wang et al. 2007, Gilbank \& Balogh 2008; Kimm et al. 2009). 
This issue has been addressed by several more recent SAM
in slightly different ways (e.g. Kang \& van den Bosch 2008; Font
et al. 2008; Weinmann et al. 2010; Kimm et al. 2011). G11
use a
gradual  stripping  of  the hot  gas  by
ram-pressure and  tidal 
effects, which is  physically well motivated
by SPH simulations  (e.g. McCarthy et al. 2008). It  is thus useful to
redo the  comparison between model  and observations with  the current
state-of-the-art SAM,
 and to  extend it to fainter galaxies than
previously done. We find indications that environmental effects
may still be overefficient in G11.

This paper  is organised as  follows. In section  \ref{sec:method}, we
describe the observational galaxy samples and the SAM. In
section \ref{sec:results},  we present  the results of  the comparison
between  SAM and  observations.  In  section  \ref{sec:discussion} and
\ref{sec:summary}, we discuss and  summarize our results.  We assume a
WMAP1 cosmology (Spergel et al. 2003),
following G11, and assume
$h$=0.73 for all masses, distances and absolute magnitudes throughout the paper. 
We note that this cosmology differs from the most recent results from WMAP7
by Komatsu et al. (2011). As discussed in more detail in G11,
this affects the abundance of high mass clusters in the model,
which probably causes the overprediction of small-scale clustering of low mass galaxies
in the SAM. G11 claim that the difference in cosmology does not seem to 
affect the distribution of galaxies within haloes much.

In all of what follows ``central 
galaxies'' are defined as galaxies that are the most massive galaxy 
in their group, while ``satellite galaxies'' are all other group
galaxies. Isolated galaxies which do not have detected satellites
are also called ``centrals''.

\section{Method}
\label{sec:method}
\subsection{Observations of Nearby Clusters}
\label{sec:obs}

In the following subsections we outline
 our observational
galaxy samples for the Virgo, Fornax, Coma and Perseus cluster. We also 
discuss cluster mass estimates, as they are important for the
comparison to the SAM. A
more detailed description
 of the sample selection and the photometry is
given in Appendix~\ref{app:obs}. All magnitudes are corrected for
Galactic extinction (Schlegel et al. 1998), but generally not
k-corrected. Our samples probe
  the dwarf galaxy regime at luminosities above $\sim$ $10^8$ L$_{\odot}$.

\subsubsection{The Virgo cluster}

Our Virgo sample is based  on the Virgo Cluster Catalog (VCC, Binggeli
et al. 1985; Binggeli et al.  1993), with membership updated by one of
us (T.L.)  in May 2008 through  new velocities given  by the NASA/IPAC
Extragalactic Database (NED), many of which were provided by the Sloan
Digital  Sky   Survey  (SDSS, Adelman-McCarthy  et al. 2007).   Galaxies  with  $v_{\rm   helio}  \ge
3500\,$km/s were  excluded; the remaining galaxies  have velocities of
$-730 \le v_{\rm helio} \le 2990\,$km/s.

Total $r$-band magnitudes and colours from $ugriz$-bands were measured
by Lisker et al. (2003), Janz  \& Lisker (2009), and Meyer et al.\ (in
prep.) on SDSS  data release 5 images (Adelman-McCarthy  et al. 2007).
For a small fraction of  the sample, $r$-magnitudes were obtained from
the VCC $B$-magnitudes  through a type-dependent $B-r$ transformation.
We  use a  distance  modulus  of $m-M=31.09$\,mag  (Mei  et al.  2007;
Blakeslee   et  al.   2009)   for  all   galaxies,  corresponding   to
$d=16.5$\,Mpc, with an uncertainty of 
1.4 Mpc, i.e. 0.15 mag (Blakeslee et al. 2009).  Our  final working sample only  includes galaxies with
$M_r\le-15.2$\,mag, since our sample can be considered complete in $r$
down to this  value (see Appendix).  Only galaxies out to  a projected clustercentric
distance (calculated  from the central  galaxy M\,87) of 1.5  Mpc, the
spatial completeness of the VCC, are included. Due to this limitation, we omit the southern
subcluster around M\,49.

 McLaughlin et al. (1999) estimate  a  virial  radius  of  1.5\,Mpc for the
Virgo cluster (adapted  to  WMAP1  cosmology),
corresponding to a  mass of  $4.0 \cdot 10^{14}  M_\odot$.
 Together  with the
finding  of Schindler  et al.  (1999)  that the  M\,87 subcluster  is
$\sim2.4$  times more massive  than the  M\,49 subcluster,  this would
yield a mass  of $2.8 \cdot 10^{14} M_\odot$  for the M\,87 subcluster
(neglecting the  M\,86 subcluster contribution), which is the relevant
mass for our study. Schindler et al. (1999) estimate a mass of $1.7 \cdot 10^{14} M_\odot$
within 1.2\,Mpc for  the M\,87 subcluster, and Urban et al. (2011) estimate a virial mass of $1.4 \cdot 10^{14} M_\odot$.
All of these estimates are in agreement
with  B\"{o}hringer  et  al.  (1994),   who  quote  a  mass  range  of
$1.2-5.0\cdot  10^{14}  M_\odot$ for  the  M\,87  subcluster within  a
radius  of  1.5\,Mpc  (values  scaled  to  our  cosmology).  
We
thus  estimate the  Virgo mass  relevant here  to  be 
$1.4$--$4 \cdot 10^{14} M_\odot$.

\subsubsection{The Fornax cluster}

Our  Fornax  sample is  based  on  the  Fornax Cluster  Catalog  (FCC,
Ferguson 1989).  By applying the  type-dependent $B-r$ transformations
derived from the  Virgo galaxies to the FCC  $B$-magnitudes, we obtain
$r$-magnitude estimates.

We use a distance modulus  of $m-M=31.51$\,mag (Blakeslee et al. 2009)
for all  galaxies, corresponding to $d=20.0$\,Mpc, with an
uncertainy of 1.6 Mpc, i.e. 0.17 mag.   Our final working
sample   only    includes   galaxies  down to the completeness limit of
$M_r=-15.9$\,mag (see Appendix).
The Fornax cluster  has two main components, with  one subcluster that
seems to  be infalling for the  first time (Drinkwater  et al. 2001).
In our analysis, only galaxies  out to a
projected clustercentric distance  (calculated from the central galaxy
NGC\,1399)  of 0.9  Mpc
  are
included, which means that we omit the Southwest subcluster.
Drinkwater  et al. derive  a mass  of  $5\pm2 \cdot 10^{13} M_\odot$
from integrating the velocity amplitude profile, and of $9 \cdot 10^{13} M_\odot$
from the projected mass virial estimator
for the main cluster.
\subsubsection{The Coma cluster}
\label{sec:coma_obs}

We use two different observational samples for the Coma cluster, which
we  find to be  consistent in their region of overlap. The  first sample  (named ``Coma''
hereafter)  is   given  by  Michard   \&  Andreon  (2008),   based  on
spectroscopic and  morphological membership criteria  and covering
the area   within 0.5 Mpc of the cluster
center.
The  second  sample  (``ComaB'')  is
constructed from  SDSS data and,  in addition to  spectroscopic member
galaxies,  involves  a  statistical   correction  for  the  number  of
contaminating background galaxies. Where SDSS redshifts are available,
only galaxies with $4000\le cz \le10\,000\,$km/s are considered.

Total $r$-band magnitudes and  colours from $ugriz$-bands are provided
through Petrosian magnitudes  of SDSS data release 7  (Abazajian et al
2009).   We use  a  distance modulus  of  $m-M=35.00$\,mag (Carter  et
al.  2008) for  all  galaxies, corresponding  to $d=100.0$\,Mpc.
From the table of Carter et al. (2008), we estimate the uncertainty
in absolute magnitude to be $\sim$ 0.2 mag. We have
checked that this uncertainty (which is the same for Perseus) has virtually
no impact on our results.   The
``Coma'' sample  is considered  complete down to  $M_r=-16.7$\,mag 
(see Appendix) and
out  to a projected  clustercentric distance  of $0.5\,$Mpc,  with the
cluster center being defined as midway between the two massive central
ellipticals NGC\,4874 and NGC\,4889. The ``ComaB'' sample includes, by
construction,  all  galaxies  down  to  $M_r=-16.7$\,mag  and  out  to
$4.2$\,Mpc.

The best-fit  Navarro, Frenk \& White (1997, hereafter NFW) profile of  {\L}okas \& Mamon
(2003) gives  a virial  radius of 2.8\,Mpc for the Coma cluster,
  enclosing a mass  of $1.3
\cdot  10^{15} M_\odot$  (using WMAP1  cosmology). This  is compatible
with the values of Briel et al.  (1992), who find a mass of $1.3 \cdot
10^{15} M_\odot$  within 3.4\,Mpc. It  also appears in  agreement with
Kenn \&  Gunn (1982), who find  a mass of $2.0  \cdot 10^{15} M_\odot$
within  the much larger  radius of  5\,Mpc (all  values scaled  to our
cosmology).\footnote{ In  their introduction, Carlberg et al.  (1997) quote a
  much larger mass value from Kent \& Gunn (1982), but they mistakenly
  multiplied their value with $h_{50}$ instead of dividing by it.  }

\subsubsection{The Perseus cluster}

Our Perseus sample  is constructed from the SDSS  in a similar 
way as the ComaB
sample,  involving   a  statistical  correction  for   the  number  of
contaminating background  galaxies. No  SDSS spectra are  available in
this  region.    We  use  a  distance   modulus  of  $m-M=34.29$\,mag,
corresponding to  $d=72.3$\,Mpc (Struble \&
Rood  1999, using  NED). 
From the velocity distribution of Perseus galaxies shown in Fig. 1 of Kent \& Sargent (1983), it is obvious that the systemic cluster velocity cannot differ by more than $ \sim$ 500 km/s from the
 Struble \& Rood (1999) value used by NED to compute the cluster distance. 
This corresponds to an uncertainty of 0.2 mag in the absolute magnitudes.
 The  sample includes,  by  construction, all
galaxies  down  to  $M_r=-16.7$\,mag.   The spatial  coverage  in  the
cluster outskirts is  incomplete, requiring a statistical completeness
correction in  the following analysis.
 Galaxies are  considered out  to a
projected clustercentric distance  (calculated from the central galaxy
NGC\,1275) of $3.8$\,Mpc.

The most up-to-date mass estimate for the Perseus cluster is
given by Simionescu et al. (2011) as  $6.7  \cdot  10^{14} M_\odot$.
A larger mass was found by Kent \&
Sargent (1983),  exceeding $10^{15} M_\odot$, yet Eyles  et al. (1991)
remarked  that  this mass,  inferred  from  the  galaxy distribution,  is
clearly  too high as  compared to  the mass  derived from  X-ray data.
Eyles  et al.  (1991) report  a mass  of $3.4  \cdot  10^{14} M_\odot$
within 0.9\,Mpc, which is the  extent of their data. 
If we extrapolate the mass profiles for the dark matter\footnote{Since
 only their X-ray  profile is given explicitly, but not
their dark matter profile,  we approximate the latter by $\rho(r)_{\rm
  DM}=   1.0  \cdot  10^{14}   M_\odot/{\rm  Mpc^3}\cdot   h^2_{50}  /
((r/r_s)^{2}    \cdot     (1+(r/r_s)^{2}))$,    with    $r_s=0.6\,{\rm
  Mpc}/h_{50}$.  This  agrees  with  the  profile shown  in  Eyles  et
al.  (1991) to  within few  percent outside  of the  cluster  core.} and the 
X-ray gas component as given in Eyles et al. (1991, their Fig. 7) until
the
total enclosed  mass and volume  correspond to 200 times  the critical
density, we obtain a mass of  $4.6 \cdot 10^{14} M_\odot$.

\subsection{The Yang et al. group catalogue}
We use the  publicly available SDSS DR4  group catalogue of 
Yang et al. (2007)\footnote{The Yang et al. group catalogue can
be downloaded from \texttt{http://www.astro.umass.edu/$\sim$xhyang/Group.html}}
for 
Fig.  \ref{fig:disto} in this  paper, where we compare the colours
of low mass/faint satellite and central galaxies in the SAM and the SDSS.
 This  group catalogue  has been
constructed  by  applying  the  halo-based  group finder  of  Yang  et
al. (2005) to the New York University Value-Added Galaxy Catalogue (NYU-VAGC; see
Blanton  et al.   2005a).   From  this catalogue,  Yang  et al.  (2007)
selected all  galaxies in  the main galaxy  sample with  an
extinction-corrected apparent magnitude  brighter than
$m_{r}=18$ and a redshift
in the range  $0.01<z<0.20$, with a redshift  completeness $C_{z} >
0.7$.  We refer
the reader to Yang et al.  (2007) for a more detailed description.

\subsection{The Semi-Analytic Model}
Semi-analytic  models  apply simple  analytical  recipes to  dark 
matter merger
trees generated  analytically or from N-body simulations,  in order to
model   the   formation   and   evolution  of   galaxies   over   time
(e.g. Kauffmann et al. 1993; Cole  et al. 2000).  Here, we use the SAM
by  G11,  applied  to  the subhalo merger trees of the MS-II simulation.

 The  MS-II simulation is  a dark
matter-only simulation, which was carried out in  a periodic box of side
137 Mpc, with  a particle mass of $9.45  \cdot 10^{6}
M_{\odot}$. This corresponds to
a  mass  resolution 125  times higher than  the
Millennium  Simulation (Springel et  al. 2005),  onto which many 
of the recent SAMs have been built (e.g.  Bower  et al.  2006, Croton et
al.  2006, De  Lucia \&  Blaizot 2007, Font  et al.  2009,  Neistein \&
Weinmann 2010).  
G11 based their model on De Lucia \& Blaizot (2007), but
 modified it in several  aspects. In particular, the
efficiency  of  SN  feedback for low mass galaxies was  increased
considerably in  order to  fit the  low mass end  of the  stellar mass
function. Also,  the prescriptions used to calculate  galaxy sizes
 and
environmental effects  have been modified,  and the efficiency  of AGN
feedback was increased. As the prescriptions for environmental effects
are  most important for  this work,  we describe  them in  more detail
below.

\subsubsection{Environmental effects}
\label{sec:env_SAM}
In the MS-II simulation, Friend-of-Friend (\fof hereafter) haloes are defined 
by linking particles with separation less than 0.2 of
the mean value (Davies et al. 1985). Within these \fof haloes, subhaloes
are identified using the \subfind algorithm (Springel et al. 2001). This means that 
galaxies within groups and clusters are associated with a dark matter subhalo
and follow its orbit, until the subhalo mass
falls below the mass of the galaxy, where an analytical prescription 
for the orbit takes over (see G11).

The treatment  of environmental effects  in G11 differs
notably from  most previous SAMs, like De
Lucia \& Blaizot (2007).  We  refer the interested reader to 
G11
for  more  details, and  briefly  summarize  the most  important
changes with respect to De
Lucia \& Blaizot (2007)
below.   First of  all,  the definition  of  a ``satellite''  galaxy
-- which is here
 a galaxy being affected by environment -- has changed. In
De   Lucia  \&   Blaizot   (2007),  all   galaxies   belonging  to   a
\fof group were  considered satellites,  while in  G11,
  this is  only the case  for galaxies which  reside inside
$R_{\rm  vir}$\footnote{G11 define $R_{\rm  vir}$ as the radius of the largest sphere with 
the potential minimum of the FoF group as its center, and a mean overdensity
exceeding 200 times the critical value.}.
 This reduced  the  total  number of  
satellites at $z=0$ in G11 by  a factor of two.   
Second, going back to Kauffmann et al. (1993), the standard
approach in most SAMs has been to immediately strip the extended gas
halo when a galaxy became a satellite. In G11, this has
been changed. The stripping of the hot
gas  halo  is  calculated based  on  an  estimate  on the  tidal  and
ram-pressure  forces  experienced  by   the  satellite, given its orbit 
within the \fof halo, and  is  thus
slower. The prescription is comparable to the one used in 
Kimm et al. (2011), and more detailed than the ones
used in Font et al. (2008) and Weinmann et al. (2010).
 Finally, no
prescription for  satellite disruption  had been 
included in  most previous SAMs like
De Lucia \& Blaizot (2007).  In G11, it is
assumed that  ``orphan'' galaxies  (i.e. galaxies without  a remaining
dark matter subhalo)  are disrupted entirely as soon  as the host halo
density  at  pericenter exceeds  the  average  baryon  density in  the
satellite. The  stars of  the disrupted galaxy  are then added  to the
intracluster medium. Partial disruption is not accounted for, and
also tidal heating is neglected. The latter leads to an expansion of the
system, which would facilitate tidal stripping.

\subsubsection{SAM subsamples}

Below, we describe the samples of  SAM clusters we use to compare with
observations. Samples V, F, C, P are used to compare with the
Virgo, Fornax, Coma and Perseus cluster respectively.  All results are obtained for
averaging between  3 sightlines  (along the  $x$, $y$, and  $z$ axis  of the
simulation box). The cluster center  is defined as the location
of the central galaxy in the \fof  group, which is placed at the potential
minimum of the \fof group by the SAM. For sample C, P and V, we
consider a galaxy to be a cluster  member  if  it  is  within +/-  2000  km/s
line-of-sight velocity from  the cluster center.
 This  is lower than the observational cuts for the Coma sample,
and also appears relatively narrow as compared to the velocity
  dispersion of galaxies in the center of Perseus when not corrected for 
possible velocity anisotropies, which is $\sim1300$\,km/s (Kent
\&  Sargent 1983).  However,  we  have checked  that  using 3000  km/s
instead  for sample  C and P  makes  no difference  to  any of  our
results, as extremely  few galaxies lie between 2000  and 3000 km/s
in  the  SAM. Note that our selection roughly mimics 
the method used to select cluster members for Virgo and for the central
Coma sample. It specifically includes interlopers which 
do not belong to the actual \fof group, with an interloper fraction of below 5 \% in the 
cluster center, reaching 20 \% at 1 Mpc. For all other samples, the observational selection is more
complex and cannot be easily mimicked in the SAM. It is possible that
completeness and contamination of our SAM samples is slightly different from 
the observational samples, which we have to bear in
mind in what follows.

   An outer radial limit of 1 Mpc or 1.5 Mpc in projected
cluster-centric  distance  has been applied in  part of the following analysis,  as
indicated there. We  use absolute magnitudes  from the SAM, 
which are computed using the  Bruzual \& Charlot (2003) stellar evolutionary synthesis
  models and assuming a model for dust obscuration, as described in G11.

\begin{itemize}

\item {\bf Cluster Sample V}\\
Sample V encompasses 12 SAM clusters with masses 
$1.4 \cdot 10^{14} M_{\odot}$ -  $4 \cdot 10^{14} M_{\odot}$, with
a median mass of $ 1.7 \cdot 10^{14} M_{\odot}$.

\item {\bf Cluster Sample F} \\
Sample F contains 37 SAM clusters with masses between 
$5 \cdot 10^{13} M_{\odot}$ and $9 \cdot 10^{13} M_{\odot}$, 
with a median mass of  $5.6 \cdot 10^{13} M_{\odot}$.  A galaxy is assumed
to be a F cluster galaxy if it is within +/- 1000 km/s
line-of-sight velocity from the cluster center.

\item{\bf Cluster sample C}\\
The cluster sample C only includes the most massive cluster
in the model, with a virial mass of $10^{15} M_{\odot}$.

\item {\bf Cluster Sample P}\\
The cluster sample P includes the three most massive
clusters in this model, with virial masses of $10^{15} M_{\odot}$, 
 $4.7 \cdot 10^{14} M_{\odot}$ and  $4.25 \cdot 10^{14} M_{\odot}$.

\end{itemize}

The general properties of the SAM clusters are compared
to the observed clusters in Fig. \ref{fig:abn}.
\section{Results}
\label{sec:results}
\subsection{Radial number densities and abundances}

\begin{figure}
\centerline{\psfig{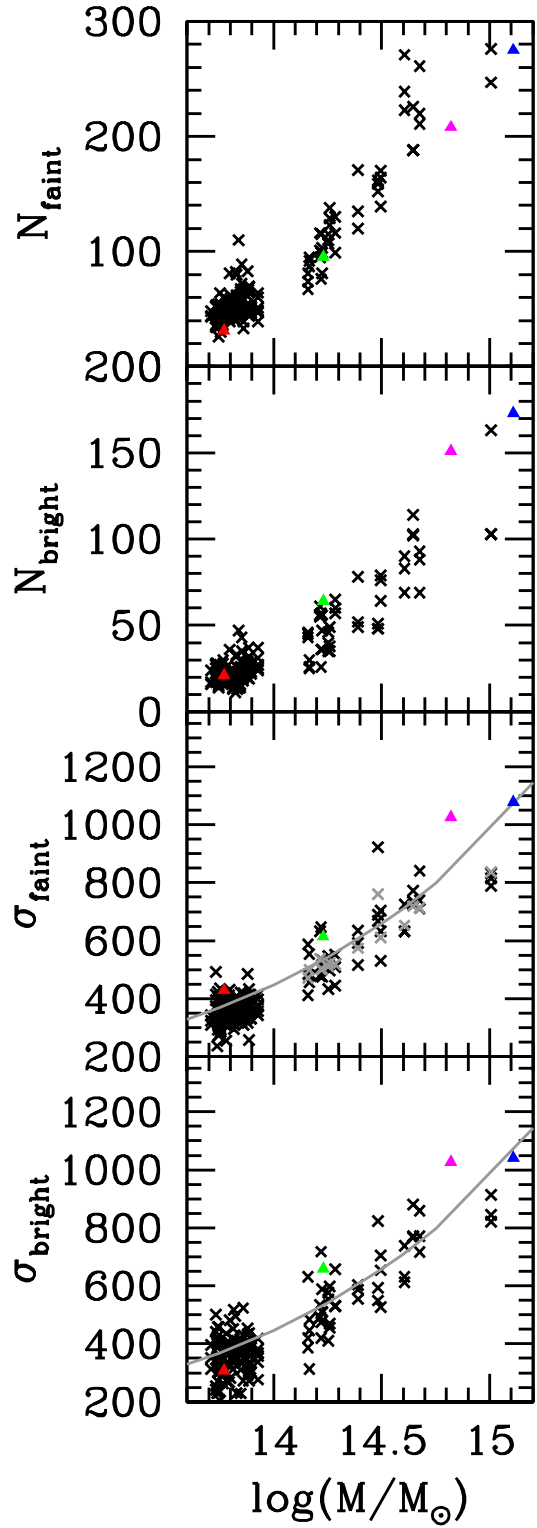}}
\caption{The galaxy abundances and velocity dispersions
 of the observed clusters (triangles; blue for Coma, magenta
for Perseus, green for Virgo and red for Fornax) and
SAM clusters (black crosses), 
measured
within 1 Mpc (0.9 Mpc for Fornax and sample F). Bright and
faint galaxies are defined according to the text. 
Grey points show the 1-D velocity 
dispersion of the SAM dark matter halo, grey lines show the fits to the 
dark matter velocity dispersion measured by Evrard et al. (2008). Here, 
we use mass estimates of $1.7 \cdot 10^{14}  M_{\odot}$, 
$6 \cdot 10^{13}  M_{\odot}$,  $1.3 \cdot 10^{15}  M_{\odot}$
and  $6.7 \cdot 10^{14}  M_{\odot}$ for Virgo, Fornax, Coma and Perseus
respectively (see sec. \ref{sec:obs}).}
\label{fig:abn}
\end{figure}

\begin{figure}
\centerline{\psfig{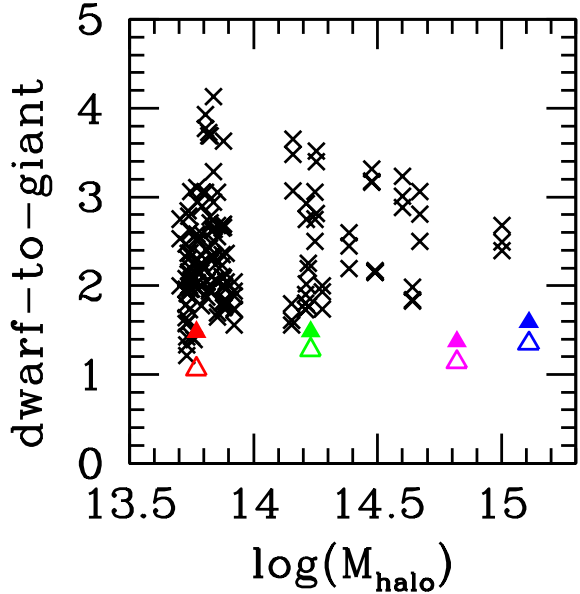}}
\caption{The ratio of the faint to bright galaxies, 
or ``dwarf-to-giant ratio''. We define faint as
 $-16.7>M_r>-19$, and
bright as $M_r<-19$. Green, blue, red and magenta filled 
triangles show
results for Virgo, ComaB, Fornax and Perseus respectively,  within
1 Mpc (0.9 Mpc for Fornax). Black crosses
show results for all SAM clusters in samples C, P, V and F within
1 Mpc (0.9 Mpc for F). Empty triangles show results for
Virgo, Coma, Fornax and Perseus within 0.5 Mpc. Masses 
of observed clusters
are as in Fig. \ref{fig:abn}.}
\label{fig:ratio}
\end{figure}

\begin{figure}
\centerline{\psfig{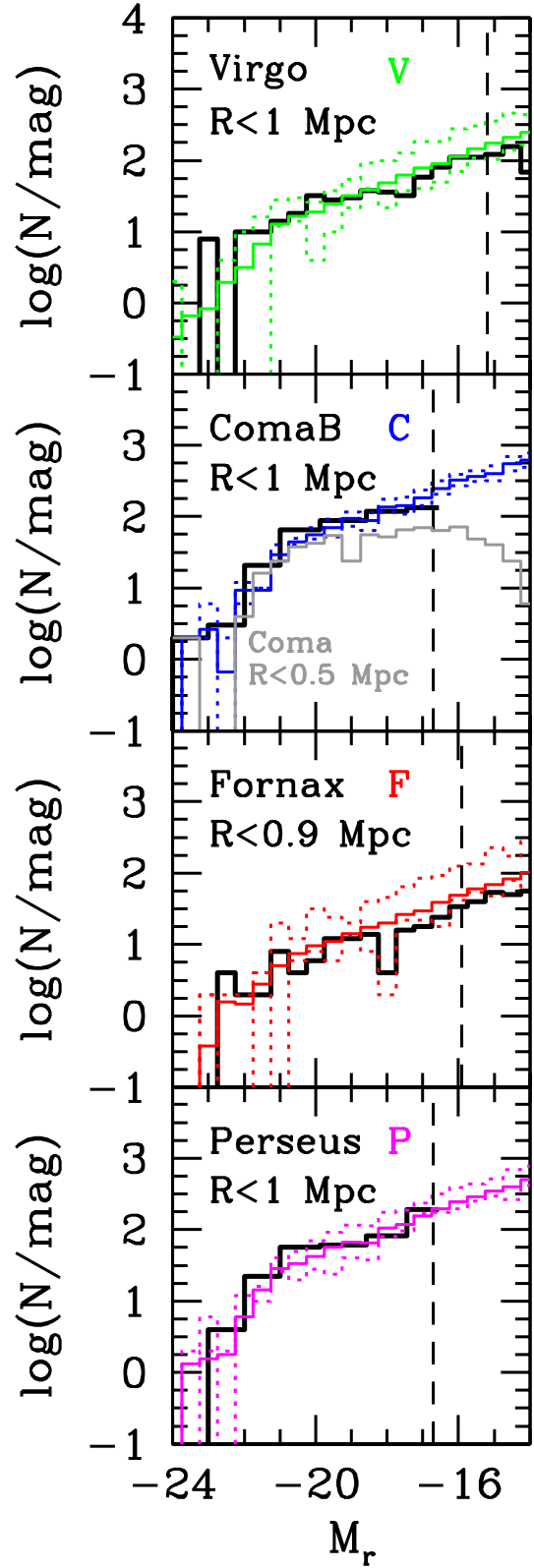}}
\caption{The LF in Virgo, ComaB, Fornax and Perseus within 1 Mpc (0.9 for Fornax), shown as thick black histograms, compared to 
the average LF in the SAM samples with the same radial cut, shown
as coloured histograms. The dotted coloured histograms show the luminosity
functions of the richest and poorest cluster in each SAM sample. For Coma, the
grey histogram shows the luminosity function in the central 0.5 Mpc.
 The dashed vertical lines indicate
the observational completeness limits.}
\label{fig:LF}
\end{figure}

In Fig. \ref{fig:abn} we show the galaxy abundances and 
velocity dispersions within 1 Mpc (0.9 Mpc for Fornax and sample F)
 of the observed
clusters and the corresponding clusters in the four SAM comparison samples, 
for both bright  ($M_r<-19$) and faint ($-16.7>M_r>-19$) galaxies. The
velocity dispersion for Perseus is taken from Fadda et al. (1996), 
where no split according to luminosity is made. Note that this velocity dispersion has been
obtained taking into account possible velocity anisotropies, which was not done in the SAM.
If no such 
modelling is included, like in Kent \& Sargent (1983), the Perseus velocity dispersion is much 
higher, at $\sim$ 1300 km/s, in potential disagreement with the SAM.
For Fornax, we take the velocity dispersion
 from Drinkwater et al. (2001) for their entire sample (including
the southwestern subcluster)
with a division between bright and faint at $M_{B}$=-16. The velocity
dispersions from Coma and Virgo are estimated from the rms of
the available velocities, without clipping. 
The properties of the observed clusters are in  
good agreement with the SAM clusters. This
is encouraging: 
Fornax and Virgo do not seem to be atypical compared to simulated 
clusters, despite indications for a relatively unrelaxed state 
(e.g. Aguerri et al. 2005; Drinkwater et al. 2001).
 Fornax has a relatively low number
of faint member galaxies compared to the SAM, but is within
the scatter of similar mass SAM clusters. Both
observed and SAM clusters are in reasonable agreement with estimates
of the 1-dimensional
velocity dispersion of dark matter particles as measured from 
 simulations by
Evrard et al. (2008, grey line in the bottom two panels) and with the same quantity measured in
MS-II for our cluster sample (grey crosses in the third panel
from top).
We thus do not find any indication for a significant 
velocity bias between
dark matter and galaxies here.
The only clear systematic difference between SAM and observations is that the
ratio between faint and bright galaxies is too high in 
the SAM.

We highlight this result in Fig.  \ref{fig:ratio}, where we show the 
    ratio  of   faint   to  bright   galaxies 
(``dwarf-to-giant'' ratio) within 1 Mpc from the cluster center (0.9
for Fornax and sample F).
 All observed clusters 
(filled coloured triangles)
have strikingly similar dwarf-to-giant ratios of $\sim$ 1.5, while 
SAM clusters (black crosses) have a median dwarf-to-giant ratio of $\sim$ 2.5 for samples 
V, C and P and of $\sim$ 2.2 for sample F. This corresponds to 
about 50-70 \% too many faint galaxies per bright galaxy.
We show the dwarf-to-giant ratio within 0.5 Mpc of the cluster center 
for the observed clusters
as empty coloured triangles. They are again all very similar, and slightly 
lower than for the more extended region in all clusters. This is in 
agreement with the decrease of the dwarf-to-giant ratio towards
the centers of other clusters (e.g. Sanchez-Janssen et al. 2008).
No similar decrease is found for the SAM samples (not shown here), 
where the median dwarf-to-giant ratio within 
0.5 Mpc from the center is 2.1 for sample F, and
2.7 for samples V, C, P, and has a very similar scatter as within 1 Mpc.

In Fig. \ref{fig:LF}, we show the LF of Virgo, ComaB, 
Fornax and Perseus within a projected distance 1 Mpc (0.9 for Fornax) compared to 
the average LF in the SAM samples with the same radial cut (coloured lines). For Perseus and
ComaB, each magnitude bin has been background-corrected separately, as
described in the Appendix. In grey, we show the LF of the central
region of Coma.
Dotted 
coloured lines denote the LF in the respective SAM samples with the minimum
and maximum number of galaxies brighter than the observational limit. Dashed
vertical lines indicate the observational completeness limit.

The shape of the LF is different in the SAM and in observations. 
While the SAM LF has a single slope 
from $M_r \sim -21$ to faint magnitudes, the
LF of Virgo, Fornax and the central region of Coma show a dip at intermediate
luminosities (at $M_r \sim -17.5$ for Virgo, $M_r \sim -18$
for Fornax and  $M_r \sim -19$ for Coma),
giving it the appearance of a double-Schechter function.
This is in qualitative agreement with previous observational 
results for the LF in clusters (e.g. Godwin \& Peach 1977 for Coma; Binggeli et al. 1988
 and McDonald et al. 2009 for Virgo;
  Popesso et al. 2005; Rines \& Geller 2008; Lu et al. 2009). The dip
is however not discernible in the background-corrected samples for ComaB and
Perseus, perhaps
due to the large magnitude bin size we are forced to use there. 
We check whether a dip occurs in the luminosity function of
\emph{individual} SAM clusters, and indeed find that a dip with a similar
magnitude like for Virgo occurs in $\sim$ 1/3 of SAM clusters, at different
absolute magnitudes. This may indicate that the dip is not physically significant, but
due to chance fluctuations.
For all clusters, the ratio between faint and bright galaxies is lower
in the observations than in the model,
 as mentioned above. The decrease of the ratio towards the 
cluster center can again nicely be seen for the central region 
of Coma, where the faint end of the LF is flattened.
We  also find again that Fornax has a low number of faint members,
in agreement only with  the 
poorest cluster in sample F.

We note that the median cluster mass of samples C, F, P is somewhat
lower (7, 30 and 40 \%) than the cluster masses
adopted in Fig. \ref{fig:abn} and \ref{fig:ratio} for Coma, Fornax
and Perseus respectively. Since this offset lies well
within the uncertainties in the mass of the observed clusters, and
due to the considerable scatter in the mass-richness relation, we do not scale
the galaxy content in the SAM samples to the corresponding observed
cluster masses. We have however checked the impact of such a correction
on the LFs. We find that the change is minor for sample F.
For sample C and P, the scaled version is
similar to the LF of the richest clusters (upper dotted line) and is
thus in slightly improved agreement for bright galaxies, but in 
worse agreement at the faint end.

\begin{figure*}
\centerline{\psfig{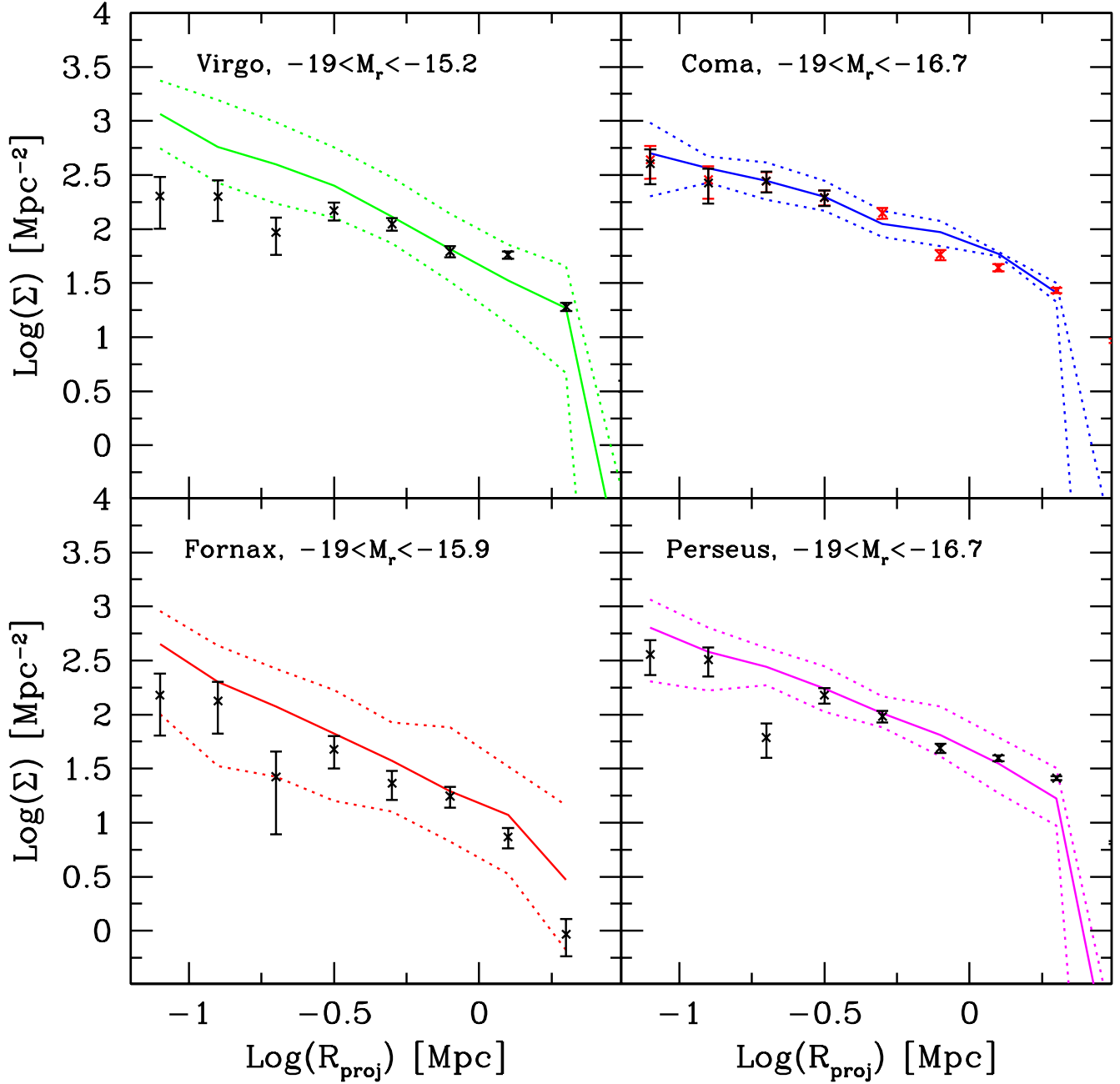}}
\caption{The average
projected surface number density profiles of faint galaxies
versus cluster-centric radius 
in SAM clusters (coloured lines) and as observed in the Virgo, Coma, Fornax
and Perseus clusters (black 
data points with errorbars). The dotted coloured lines denote the minimum and maximum density
found in all 3 projections in SAM clusters in each radial bin. Green lines
are for sample V, blue lines for sample C, red lines for F, magenta for P.  
Errorbars on the observations
are Poissonian.}
\label{fig:v1}
\end{figure*}

In Fig. \ref{fig:v1},  we show radial number density  profiles for the
Virgo,  Coma,  Fornax and  Perseus  cluster, compared to the
average profiles from samples V, C, F and P respectively. Note that the
  observational sample has a different absolute magnitude
limit for different clusters  (see Sect.~\ref{sec:obs}).
As dotted  lines, we  show the  maximum and  minimum data
points for each  bin in sample V, C, F and  P, including all three
projections. For  Coma, we  show the  results for both Coma  and ComaB
(black and red  points with errorbars, respectively), which are in excellent agreement.

We  find  that the  SAM  successfully  reproduces  the number  density
profile of nearby clusters, with  the exception of the central regions
of the Virgo  cluster, where even the 
minimum SAM  profile is substantially
higher than  the observations. This is also the case if 
we define the Virgo center as in between the M84 and the M87 galaxy, 
instead of as the M87 galaxy. 
Indications for a similar deviation may be present in the Fornax cluster. 
Again, correcting the profiles for the offset in the cluster mass
leads to slightly worse agreement for Perseus and Coma, where the
corrected profiles are similar to the maximum profile (upper dotted line).

\subsection{Red dwarf galaxies in the Virgo cluster}

\begin{figure}
\centerline{\psfig{figure=fig5_wl.epsi,width=220pt}}
\caption{The colour-magnitude relation of sample V (green squares) 
and Virgo (top panel) and of sample C (cyan squares) and Coma (bottom 
panel). Galaxies within 1.5 Mpc from the cluster centers are included, 
except for Coma, where only the central 0.5 Mpc are used. SAM
data has been downsampled to match the number of
Virgo galaxies in the same region.
Colours for Coma have been k-corrected according to the redshift
of the cluster, for consistency 
with Fig. \ref{fig:red} where not all galaxies have redshifts. 
 }
\label{fig:cmr}
\end{figure}

After having verified that the SAM reproduces the global properties of
dwarfs in clusters well, we now  examine the red fraction of dwarfs in
clusters.  For  this, we  concentrate on  the Virgo, Coma
and Perseus cluster, as no colour information is available for the Fornax cluster. 
For Coma and Perseus, we k-correct colours according
to the cluster distance 
as given in \ref{sec:obs}, since we do not have redshifts
for all individual member galaxies, using the approximation
of Chilingarian et al. (2010).

 In
Fig. \ref{fig:cmr},  we show the colour-magnitude relation 
for the  Virgo sample (top panel) and
the  Coma sample (bottom panel; note that this  only includes the
central part  of Coma, see sec. \ref{sec:coma_obs}), 
compared to samples V and C.
Galaxies in the SAM clearly
reach redder colours than in both Coma and Virgo, extending up to $g-r$
$>$ 1 at  all absolute  magnitudes.  We find that the
  very red  model  galaxies are
strongly dust-attenuated, indicating an overestimate of dust
in the SAM for some galaxies. But also the  bulk of the red sequence is
clearly redder 
  in the SAM compared to  both clusters. It is unclear why this
is the case. Galaxy colours can depend on the 
details of the stellar population models (e.g. Maraston et al. 2009), but
Conroy et al. (2009) find no difference in $g-r$ colours for evolved populations
between Maraston (2006) and Bruzual \& Charlot (2003). Also, Conroy et al. (2009) find
that variations in the initial stellar mass function (IMF) do not affect
colours significantly (except in the near-IR). The offset thus may be due to an offset 
in age or metallicity. We find that the majority
of G11 SAM cluster dwarfs have mean stellar ages in excess of 9 Gyr, clearly 
higher than the observed median stellar ages of $\sim 6$ Gyr for low mass 
cluster satellites (Pasquali et al. 2009), but in rough agreement with the observations of
Virgo dwarf ellipticals by Roediger et al. (2011).

We split galaxies into red and blue according to
\begin{equation}
\label{eq:colour}
(g-r)_{\rm cut}=0.4-0.03 \cdot (13+M_{r})
\end{equation}
(red line in Fig. \ref{fig:cmr}). This cut is chosen at a relatively blue
colour, so that Virgo still has a well-populated red sequence. To compute red fractions  for the ComaB and Perseus
sample as a function of absolute magnitude and cluster-centric distance, we background-correct the  total and red sample separately, as
described in the Appendix. For the red fractions as a function
of velocity, we only use galaxies in sample ComaB with measured
velocities. No
velocities are available for the Perseus sample.

\begin{figure*}
\centerline{\psfig{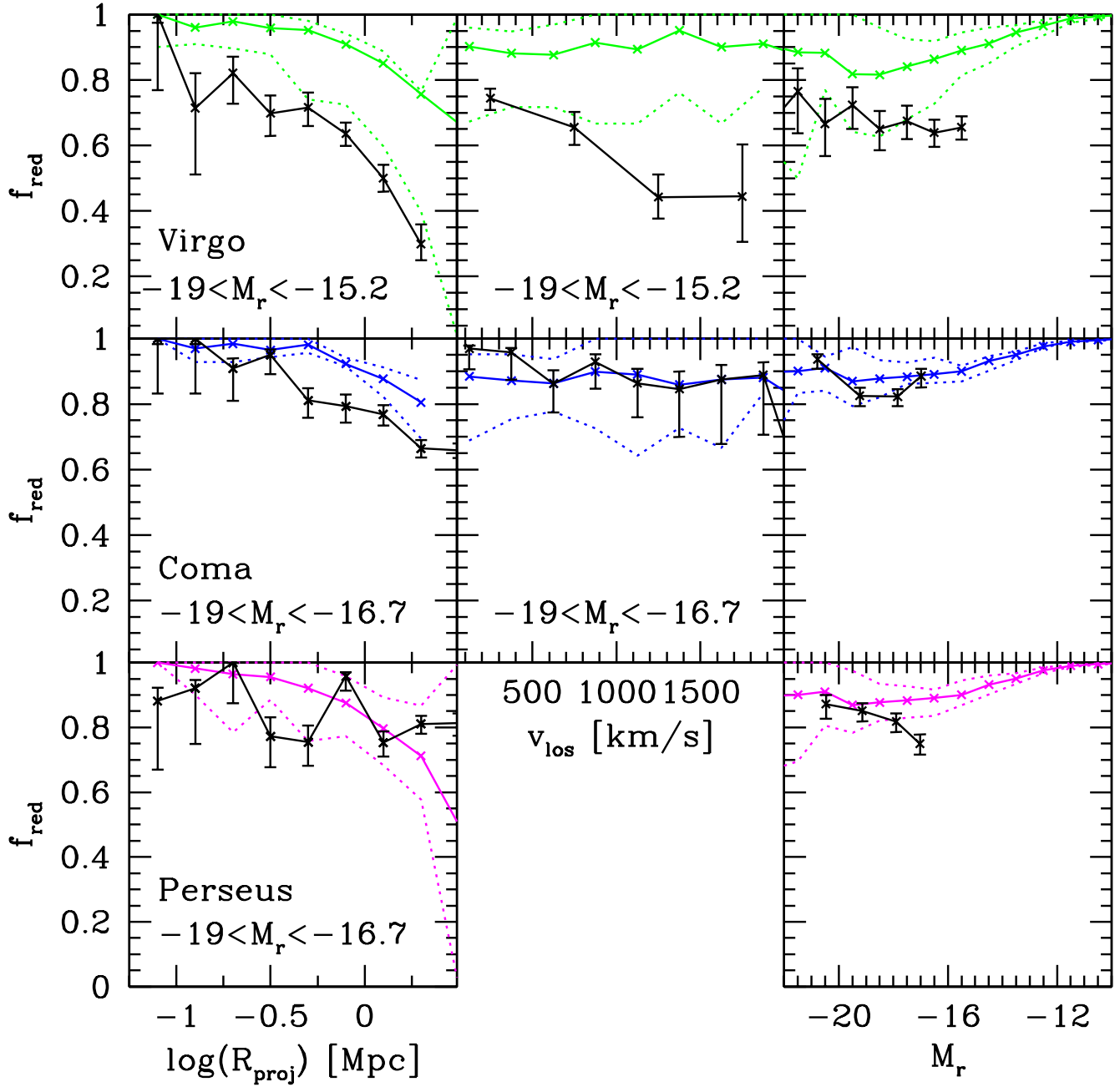}}
\caption{Red fractions  as a function of projected cluster-centric radius,
line-of-sight velocity with respect to the cluster center, and absolute magnitude. 
Observations are shown as black lines with errorbars,
results from the SAM as coloured lines. Dotted lines show the minimum
and maximum red fractions for the clusters in the SAM samples, including
all projections. Results for Virgo and
sample V are shown in the top panels, for ComaB and sample C in the middle
panels, for Perseus and sample P in the bottom panels. Errorbars denote the 
confidence
interval estimates for c=0.68 from quantiles of the beta distribution, 
which is the method recommended to calculate error on fractions by Cameron 
(2010). For the middle and right hand plot, only galaxies out to 1.5 Mpc projected
radius are used. Velocities are available for the majority of the 
galaxies in the ComaB and Virgo sample.}
\label{fig:red}
\end{figure*}

 In Fig. \ref{fig:red}, we
show the fraction of red galaxies   as  a  function  of  projected
cluster-centric radius  (left-hand panels, only for  faint galaxies, as
indicated), the line-of-sight velocity with respect to the cluster
center  (middle panels, only for faint
galaxies as indicated) and  the absolute magnitude (right hand panels)
for Virgo, ComaB and Perseus and their corresponding SAM samples. Again,
dotted lines denote the absolute minimum and maximum of the red
fraction in each bins for all SAM clusters considered, in all
three projections.

  We  find that the fraction of red galaxies
in the SAM is too high  compared to Virgo. The offset
seems  to be  roughly independent  of cluster-centric  radius,  but is
strongly  increasing with increasing  line-of-sight velocity  and also
increasing towards fainter absolute  magnitudes.  On the other hand, the
red fractions of Coma and Perseus are in 
good agreement with the model, 
with some indication that the faintest Perseus galaxies and the galaxies
at the outskirts of Coma have a low
red fraction compared to the model.

\section{Discussion}
\label{sec:discussion}
\begin{figure*}
\centerline{\psfig{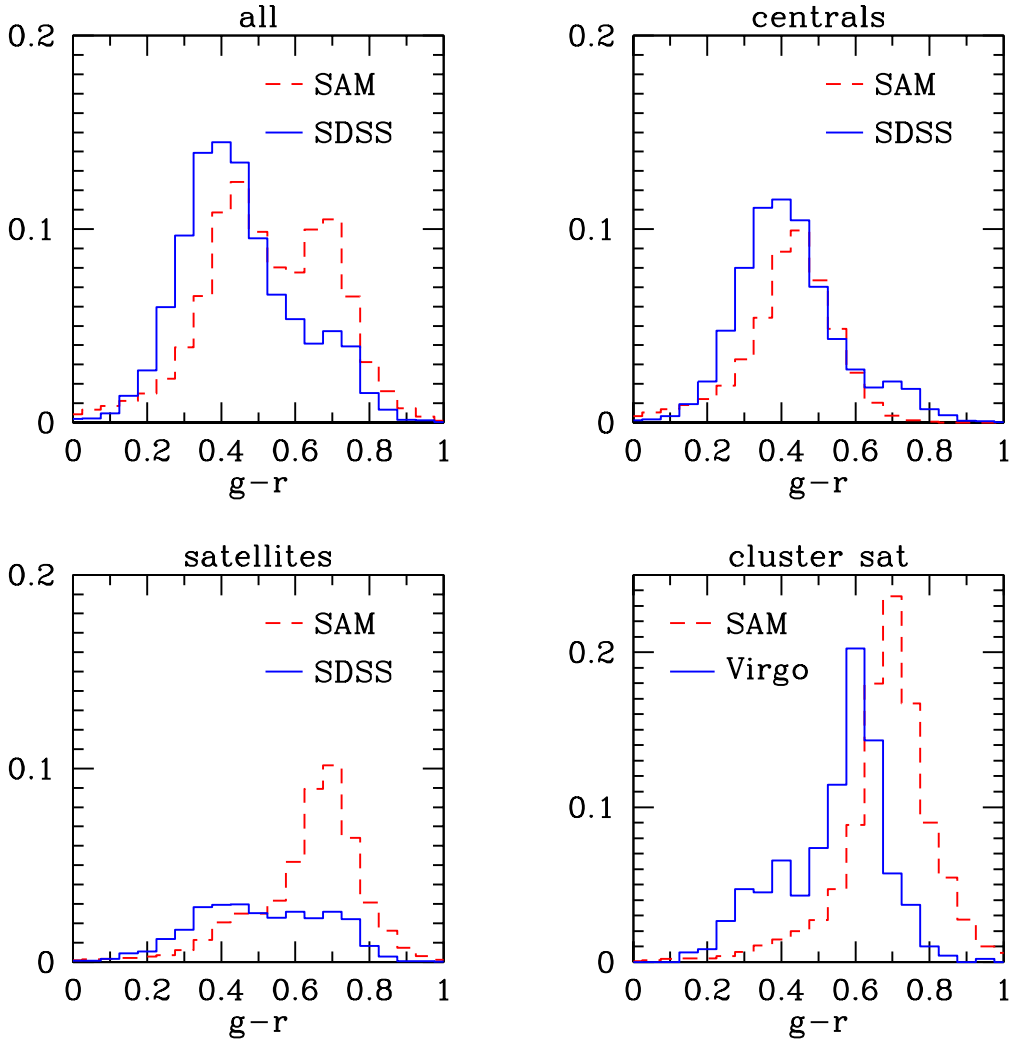}}
\caption{The colour distribution of galaxies with $-15.5>M_r>-19$  in the 
SDSS DR4 (blue solid
histograms) and the SAM 
(red dashed histograms). 
The top left panel shows the distribution of all galaxies, 
the top right panel for centrals, the bottom left panel
for satellites and the bottom right panel for Virgo and 
for haloes with masses above  $\log(M/M_{\odot})=14$ (for SAM).  Satellites
in the SAM are defined as all galaxies which are not the central 
galaxy in their \fof group (i.e. including galaxies residing
outside of $R_{\rm vir}$ of their parent \fof group).
In the SDSS, galaxy types are assigned according to the Yang
et al. (2007) SDSS DR4 group catalogue, and galaxies are weighted
according to the inverse of the maximum volume out to which they
can be observed. For Virgo, all galaxies within 1 Mpc from the center
are used, without any weighting. The histograms have been normalized
to the total number of galaxies (for SAM and Virgo) and for
the sum of all galaxy weights (for the SDSS).}
\label{fig:disto}
\end{figure*}

The  high-resolution   MS-II  simulation provides us with the  detailed
formation and  merger history
of small  mass haloes, and is thus a useful tool
for studying the evolution of  dwarf galaxies in clusters. 
If subhaloes, as well as orphan galaxies
  (without remaining subhaloes), can be mapped to dwarf
  galaxies of a given luminosity, then these model galaxies can be compared to the observed
  dwarf galaxy subpopulations  (like   BCDs,  dEs with and without
disk components) through their kinematic and spatial distribution.
In this way, one may gain new insight in the origin of those
subpopulations.  To be  able to do this,  one needs a technique to populate  the MS-II dark matter  subhaloes with
realistic  galaxies. The G11 SAM is  an obvious
choice,  as  it  is  the   first  SAM  applied  to  the MS-II
simulation.
 However, this model suffers from some
limitations. While  G11 reproduce 
the  stellar  mass  function measured in the SDSS
by Li \& White (2009)
to within 30 - 40 \% down to $\log(M_{\rm star}/M_\odot) \sim 8$ and to within $\sim$
 20 \% at $9.5<\log(M_{\rm star}/M_\odot)<11$,
the colour
distribution of galaxies with $\log(M_{\rm star}/M_\odot)<9.5$ is not matched
well, with the model displaying a clear red peak that is not
observed in the SDSS. G11
  explain this with low  mass galaxies forming  too early in
the  model and  being too  passive today  \emph{in general}.   This is
similar  to  the  claim of  Fontanot  et  al.  (2009), who
 find that several current SAMs have a fundamental problem, in 
that
galaxies with masses $9<\log(M_{\rm star}/M_{\odot})<11$ form too early and have
too little star formation today.

Given that the stellar mass function is reproduced reasonably well, 
it is not surprising that G11 predict
 a stellar mass to halo mass relation for central galaxies in agreement 
with predictions from 
 abundance matching by Guo et al. (2010)  to within $\sim$ 0.2 dex down to  $\log(M_{\rm star}/M_\odot)
\sim 7$. 
However, the match for cluster satellites may be less good due to the 
potential problems with the early history of low mass galaxies, and the fact that 
satellites in massive clusters tend to form 
at higher redshift than field galaxies (e.g. Tully et al. 2002; Neistein et al. 2011).
 It is therefore necessary to compare the general
properties of cluster  galaxies in the SAM and  in observations before
more detailed studies  are undertaken. We discuss below the results we
have  obtained  in our  comparison.

\subsection{Abundances and number density profiles of dwarf galaxies}

\subsubsection{Success in reproducing general properties}
It has been found in previous studies (e.g. Weinmann et al. 2006b; Liu
et al. 2010) that SAMs like Croton et al. (2006)
or Bower et al. (2006) strongly overpredicted
the number  of satellites  in groups and  clusters. G11
claim that their improved model reproduces the number density profiles
of galaxies with masses above $1.2 \cdot 10^{10} M_{\odot}$ in massive
clusters. In this  study, we  find that  also the
number  density  profile  of  faint  galaxies  (in  a  fixed  absolute
magnitude range) in massive clusters is well reproduced
by G11. 
The agreement can partially be fortuitous, as the too
red  colours  of   SAM dwarfs  indicate  that they will
be too faint at given stellar mass, and there might still be    an
overproduction of dwarfs at given stellar mass. We
do not use stellar mass estimates in our current study as
(i) published SDSS stellar masses for low mass members
of Virgo are unreliable due to the problems with sky subtraction
(see Appendix) and (ii) no SDSS data are available for the Fornax
cluster.

The improvement with respect to previous comparisons regarding the abundance 
of cluster satellites is mainly due to the changes in the SAM, probably
the inclusion of satellite disruption, or the more efficient
stellar feedback. We
have checked this by comparing the conditional stellar mass functions in groups 
(see Liu et al. 2010),
in G11 and De Lucia \& Blaizot (2007), and found that they
are lower in G11 and thus closer
to the observational results by Liu et al. (2010) than has been the case for
the model of De Lucia \& Blaizot (2007).

The flattening in the number density profile towards the center of Virgo 
(and perhaps also Fornax) is
interesting, since no comparable mass SAM cluster shows a similar feature.
The most obvious explanation may be that Virgo is a dynamically young cluster.
However, it is then curious that no cluster with a similar profile shows up in the
MS-II simulation box, as about 1/3 of observed clusters are found to fall into the same class 
as Virgo according to several cluster classification schemes
(Rood \& Sastry 1971; Bahcall 1977). However, of course it is possible that Virgo is 
still a special case within this class.

In all, the  deviations in the general properties of cluster
dwarf galaxies between the SAM and observations are
relatively small, which shows that the SAM can be used to
  predict in which subhaloes [or on which orphan
  trajectories] galaxies of a  given absolute magnitude should 
reside. However, some puzzling differences remain.

\subsubsection{The dwarf-to-giant ratio: An indication for
insufficient tidal disruption?}
\label{sec:tidal}

We find that the
 dwarf-to-giant ratio in all SAM clusters is high compared to 
observations (Fig. \ref{fig:ratio}). This
 problem seems confined
to clusters, as the dwarf-to-giant ratio seen in the general $r$-band LF is, 
if anything, too low in the SAM (see Fig. 8 of G11). 
Also, there is a small, but 
consistent
decrease in the dwarf-to-giant ratios towards the center of all 
clusters which is not found in the SAM. This
seems to be
 in
agreement with  the decrease in  the dwarf-to-giant ratio  towards the
centers  of other  clusters  (Pracy et  al.  2004; Sanchez-Janssen  et
al. 2008; Barkhouse et al. 2009), and the higher velocity dispersion
of dwarfs compared to giants in Fornax (Drinkwater et al. 2001) 
and Virgo (Conselice et al. 2001).

It is clear that the dwarf-to-giant ratio
in the SAM could be decreased
by either increasing the disruption efficiency for faint galaxies 
(possibly especially for slow galaxies in the cluster core), or
by not forming them in the first place.
Decreasing galaxy formation efficiency  for low mass haloes
at high redshift would be an attractive
solution, since it would also decrease stellar ages, and make colours bluer. 
However, it is not clear what could be the physical cause for such 
a decreased efficiency (e.g. Fontanot et al. 2009). Enhanced tidal disruption 
may be easier to justify. Diemand et al. (2004) point out that tidal disruption
should be especially efficient for slow subhaloes. This would 
explain the broader spatial and velocity distribution for dark matter
subhaloes compared to the dark matter particles 
(e.g. Ghigna et al. 2000; Gao et al. 2004; Diemand
et al. 2004).
 Therefore, strong tidal disruption for faint galaxies would not only lead to a lower dwarf-to-giant ratio, but also
a broader spatial and velocity distribution of the affected population, 
leading to a decrease of the dwarf-to-giant ratio towards the
cluster center.
We note that alternative explanations have
been suggested for a low central concentration
and high velocity dispersion of dwarfs in clusters.
 Drinkwater et al. (2001) and Conselice et al. (2001) suggest that dwarf galaxies in Virgo and Fornax
are an infalling population, while White (1976) predict the same
effect from energy equipartition between high and low mass
galaxies for a relaxed population.

Assuming that insufficient tidal
disruption is the main cause of the discrepancies mentioned above, 
it is useful to check the importance of this effect in G11. We find that
 disruption in 
this model is insignificant at lower 
redshift, where $< 0.1 \% $ of cluster galaxies 
are disrupted per Gyr. It is more important at higher redshift, 
with a disruption efficiency of 0.6 \% per Gyr at z=2, and
1.2 \% per Gyr at z=4 in the most massive clusters at these times. 
The increase of tidal stripping 
with redshift is in qualitative agreement with some earlier results
(Giocoli et al. 2008; Tinker \& Wetzel 2010; Weinmann et al. 2010, 
De Lucia et al. 2010,
but see also Murante et al. 2007). 
Interestingly, we find in our checks that the disruption efficiency in the SAM
is roughly independent of the stellar mass of the galaxy -- a similar
fraction of massive and dwarf galaxies in clusters is disrupted, 
in agreement with the result of Weinmann et al. (2010) for subhaloes.
This may be the result of two competing effects: 
On the one hand, at a fixed orbital radius, the tidal radius
goes roughly as $(m/M)^{1/3}$ with $m$ the satellite mass and
$M$ the host halo mass (Binney \& Tremaine 1987). At fixed cluster mass, 
lower mass galaxies will thus be stripped more. 
On the other hand,
the dynamical friction times are shorter for high mass satellites at
a given cluster mass, and
thus these galaxies reach the region where stripping is efficient
earlier.
Some SAMs (e.g. Kim et al. 2009) 
ignore the second effect, and assume that the
tidal stripping efficiency for satellites is proportional to $M/m$.
While this may help with the problems outlined above, since it will
destroy relatively more dwarf galaxies compared to massive galaxies, 
a detailed physical justification for such a scaling has yet to be established.
Other models assume the opposite dependence, with the tidal
stripping efficiency  scaling as $m/M$
(Wetzel \& White 2010), with
some indirect support from observational results (Yang et al. 2009).

We have checked whether a simple increase in the disruption efficiency
in the SAM can solve the problem, by assuming that all orphan galaxies (i.e. 
galaxies which have lost their associated dark matter halo)
are completely disrupted, following Henriques et al. (2008). 
This is the most extreme case for the current
implementation of disruption in G11 since only orphan galaxies are eligible for disruption 
in the first place. We find that this does not change the average dwarf-to-giant
ratio in SAM clusters significantly. It seems therefore that the solution 
would have to involve a different parametrization for disruption
than the one used in G11, which prefers disruption of low mass galaxies, like suggested
by Kim et al. (2009).  Maybe taking into account tidal heating, which leads to the
expansion of affected systems, and make them more vulnerable to tidal stripping, or
the inclusion of partial tidal disruption, would help.

We note that increasing the fraction of tidally
disrupted dwarfs in the SAM will not lead to 
disagreement with predictions for the intracluster light. Even if all 
dwarf galaxies in massive clusters were additionally completely disrupted, this
would only change the stellar mass in the intracluster stars
by $\sim 20 \%$ compared to G11. This is a clear improvement compared
to earlier SAMs, where disrupting all the surplus galaxies would have
led to an overprediction of the intracluster light (Liu et al. 2010). This
is due to the fact that the number of massive cluster galaxies
are better reproduced by G11 than by earlier models.

Clearly, more work is needed to understand this issue better, and in particular
to disentangle the effects of low mass galaxy formation on one
hand, and disruption 
and gas stripping in clusters on the other hand. To this aim, it would perhaps be helpful
to study the evolution of the dwarf-to-giant ratio (defined according to mass) with 
redshift both in the field and in clusters, and compare with the SAM. Previous results
suggest a significant 
evolution in the red dwarf-to-giant ratio in clusters (De Lucia et al. 2007, 
Gilbank \& Balogh 2008, Lu et al. 2010), but this does not have to hold for the total
dwarf-to-giant ratio.

\subsection{Red fractions}
 The red fractions for model clusters are in good agreement with Coma
and Perseus,
 but clearly higher  than what is observed in  the Virgo cluster 
(Fig. \ref{fig:red}).   This could have at least
four different reasons:
\begin{enumerate}
\item The  red fraction  of cluster galaxies  in the SAM  are correct.
  Virgo is  an extreme case of  a very unevolved cluster,  which is so
  rare  that it  does not  show  up in  the MS-II  simulation
  box. Indeed, Virgo is probably dynamically young, given its clumpy
X-ray distribution and inhomogeneous intracluster light (e.g. Binggeli
et al. 1987 and Aguerri et al. 2005). However, as mentioned before, 
according to Bahcall (1977), about
1/3 of all clusters should probably be similarly unrelaxed.
\item  The problem  is mainly  due to  too high  red fractions  in the
  population of \emph{central} (``field'') 
dwarf  galaxies in the SAM, coming from
  a too  early formation  of those galaxies.   The efficiency  of star
  formation quenching due  to environment in the SAM  is correct. This
  corresponds to the  interpretation given in G11 where
  they say  that the  principal reason for  too red satellites  in the
  model is that central galaxies are too passive.
\item The problem is mainly due to overefficient quenching of star 
formation in satellite galaxies. Its severity depends on the formation 
history of the cluster. If most galaxies have been satellites for a long
time, as might be the case in Coma, the problem is hidden. If there is a significant population
of galaxies which have fallen in relatively recently, as in Virgo, 
the problem becomes apparent, since such galaxies are already red in the SAM and
still blue in the observational data.
\item The solution is to remove 
the reddest $50 \%$ of dwarf galaxies in clusters. This would bring
both abundances and red fractions into better agreement: It would bring
the 80 \% red galaxies at $M_r$=-15.5 in the bluest SAM cluster down to 60 \%, in 
agreement with Virgo. This might
be achieved by increasing tidal disruption for low mass galaxies, 
as discussed in section \ref{sec:tidal}, or by inhibiting the formation 
of those galaxies at high redshift 
in the first place (see also Liu et al. 2010).

\end{enumerate}

Another explanation could be an extremely high interloper fraction in Virgo.
We however consider this to be unlikely, as the total number of member galaxies, and
their velocity dispersion, seems in good agreement with what is expected 
for the Virgo mass, calculated from the X-ray properties of the gas, e.g. 
by Schindler et al. (1999). Also, there is a clear discrepancy between the red
fraction in the SAM and in Virgo already in the central part of the
cluster, where the interloper fraction should be lowest.

\subsubsection{Comparison between the SAM and the Yang et al. group catalogue}
To find out which of the above scenarios is most likely, we  make  a  
comparison of $g-r$ colours
of central and satellite  galaxies in the SAM  and the
Yang  et  al.  (2007)  group  catalogue  based on  the  SDSS.  
About 70 \% of all satellites in this catalogue reside in groups
with masses $<10^{13}$, making
it complementary to the cluster sample we have used above. In  the
observations,  the number  of galaxies  is weighted  according  to the
inverse of the maximum volume out  to which they can be observed given
their     absolute     magnitude. Colours are k-corrected
to z=0.0.      Results    are     shown     in
Fig.  \ref{fig:disto} for galaxies with  $-15.5 < M_r  < -19$.

By splitting the population into centrals and satellites both
in the observations and in the model, we  see where the discrepancy
between the SAM and the observations
comes  from.
Clearly,  the main  problem are  the  satellite galaxies,
which are much redder and more abundant than in the observations (bottom left
panel of Fig. \ref{fig:disto}), and practically solely
responsible for the red peak in the model.  This indicates that
explanations (i) and (ii) probably do not apply -- the problem is not
only confined to the Virgo cluster, but also appears in lower mass groups, 
and the problem does not seem to originate from the colours of 
central galaxies, which are roughly correct.
We have also checked the colour distribution in the stellar mass bin
$\log(M/M_{\odot})$=8-8.5 both in the SAM and in the SDSS, and found 
similar results, with a pronounced red peak in the SAM mainly originating
from the satellite population that is missing in the SDSS.

\subsubsection{Too red colours of satellites galaxies}
The too red colours of satellite galaxies 
indicate that either environmental effects are still overefficient 
in G11 (option iii)
or that red satellites need to be destroyed or their
formation needs to be hindered (option iv). 
Both options seem viable. It is well possible 
 that the prescription for environmental effects used by G11 
is incorrect for groups, as ram-pressure stripping in those
environments is likely overestimated due to an overestimate in 
the hot gas content of groups (Bower et al. 2008; Weinmann et al. 2010). 
A potential
indication for overefficient ram-pressure stripping 
is that the deviation
between red  fractions in  SAM and Virgo  is strongest  at high
line-of-sight  velocities and for  faint galaxies (Fig. \ref{fig:red}),  where ram-pressure
stripping is  likely most efficient.
It is also possible that tidal disruption is significantly 
stronger than assumed in G11. As mentioned before, there
is no consensus in the models whether the 
 tidal stripping efficiency for a satellite with mass $m$
newly infalling into a group with mass $M$
scales roughly with $M/m$ (Kim et al. 2009), with $m/M$ (Wetzel \& White 2010), 
or is approximately independent of the ratio (Weinmann et al. 2010).
If it scales with $m/M$, tidal disruption of faint galaxies
will be important for low mass groups and can thus help 
to remove red galaxies in these environments (see also below). On the other hand, it
will then naturally be less significant in massive clusters, where
it may be needed to resolve the issues discussed in Sec. \ref{sec:tidal}.
Overefficient early galaxy formation, on the other hand, seems
a less likely explanation, as its impact should be smallest
in low mass groups (that form latest), which is
inconsistent with our results.

\subsubsection{Discrepancy in the fraction of satellites}
Another discrepancy  concerns the fraction of satellites  in the model
and the  observations.  At $8<\log(M/M_{\odot})<8.5$,  there are only
27   \%  satellites  in   the  Yang   et  al.    catalogue  (including
volume-weighting), but $\sim 50 \%$  in the SAM if satellites outside
of  $R_{\rm vir}$  are counted,  and $\sim  40 \%$  if those  are not
counted.  This means that either (a) there are too many satellite galaxies
in  the SAM,  or that (b) a significant fraction of SDSS  satellites is
not detected, or  (c) a significant fraction of SDSS satellites is misclassified
as centrals. It is possible that there is a contribution from
effect (b), as the
SDSS spectroscopic sample is incomplete due to fiber collisions, and this problem becomes
more severe in dense regions dominated by satellite galaxies
(with the incompleteness reaching around 60 \% close to cluster centers according to G11).
Point (c) seems not significant, as tests indicate that there are more centrals wrongly 
classified as satellites than the other way around (Weinmann et al. 2009).
But if the satellite galaxies that are missed in the SDSS have a similar
colour distribution as the ones that are detected, incompleteness 
will not solve the discrepancy in the colour distribution between SDSS and SAM.
Removal of low mass satellites by tidal disruption  would of course 
help resolving the problem of overabundant faint satellites. Also, note that
using a cosmology with a lower $\sigma_8$ (like WMAP7) will reduce the fraction of 
satellites compared to WMAP1 slightly (e.g. van den Bosch et al. 2007).

\subsubsection{The colours of central galaxies}

The deviation between  model and observations
for the  central galaxies is quite  small, with an offset  in the blue
peak   of   central   galaxies   of   around   0.05 for $-15.5 < M_r  < -19$, 
and a somewhat too low fraction of galaxies with $g-r$ $<$ 0.4
in the SAM (see Fig. \ref{fig:disto}, top right panel).
 This is surprising, as several recent studies
have claimed that there is a serious problem at the low mass
end in SAMs, in that low mass galaxies form too early and
are too passive and red today (e.g Fontanot et al. 2009, G11). 
It has been argued that  the same problem
manifests itself in  the missing evolution of the low  mass end of the
stellar mass  function in  the SAM (G11) and in an  overproduction of
star forming galaxies at $z \sim 4$ (Lo Faro et al. 2009). 
Similar problems
have been found in hydrodynamical simulations, where the specific
star formation rate (sSFR hereafter) of low
mass galaxies are too low by an order of magnitude (Avila-Reese et al. 2011; 
Dav\'{e} et al. 2011).
We find here that faint central galaxies are mostly blue and
star forming, with a passive fraction (defined 
as $\log({\rm  sSFR/yr})<$-11) of only 25 \%, which may be explained with 
an intermittent phase in a bursty star formation history. 
We have checked that the sSFR--stellar mass relation in G11 is similar
to that in De Lucia \& Blaizot (2007), as shown in Fontanot et al. (2009).
At low masses of $\log(M) \sim 9$, sSFR are thus lower by about 0.5 dex than those estimated by
  Salim et al. (2007), but similar to those measured
by Gilbank et al. (2011).
Colours are however maybe the better indicator to use
for low mass galaxies with possibly stochastic SF histories, as they measure star formation on longer timescales
that the usual SF indicators, and can be observed  directly.
Fontanot et al. (2009) have pointed out that
 the mismatch in the properties of low mass galaxies
could indicate a fundamental problem in our understanding of high redshift 
baryonic physics. The good match in the colours of central
galaxies found here challenges this interpretation, and indicates that the
issue has to be investigated in more detail.

\section{Summary}
\label{sec:summary}

We compare  the properties  of dwarf galaxies  in the  nearby clusters
Virgo,  Coma,   Fornax  and   Perseus  to  the   G11
SAM and find that their abundances, velocity dispersions and
 number density  profiles are reproduced well by the model.  This is
encouraging for future studies,  in which the
  spatial and dynamical distribution of  different  subpopulations of
  dwarf galaxies within a given environment may be
matched to  SAM cluster  galaxies to investigate  their origin.

We find  that Coma and Perseus resemble
 a typical SAM  cluster in several important aspects, including 
the red fraction of galaxies.
On  the other  hand, no SAM  cluster with  comparable mass
displays red  fractions of faint galaxies as low as  the Virgo  cluster.
By comparing the colours of central and satellite
galaxies in the SAM to the Yang et al. (2007) group catalogue, we show that 
the too high red fraction in the SAM clusters compared to Virgo is likely
mainly due to an overestimate of environmental effects in the model, 
possibly related to overefficient ram-pressure
stripping of the extended gas reservoir of group galaxies. This 
problem may be excerbarated by insufficient tidal disruption of low
mass galaxies in the SAM, 
which lets too many old, red dwarf galaxies survive.
  The colours of satellite galaxies are also the origin of the mismatch
in the general colour distribution of low mass galaxies between the SAM
and the SDSS. This interpretation differs from the one
given in G11, where it was argued that a too early formation of
dwarf galaxies in general is the main reason for this.

Furthermore, we find that the dwarf-to-giant ratio in SAM clusters is high
compared to observations. We argue that the most likely explanation for this
difference is that the SAM underestimates tidal disruption for faint galaxies.
 In addition, we point out that the low number of faint galaxies in 
the center of Virgo is not reproduced in the SAM. 

\section*{Acknowledgments}
We thank the anonymous referee for useful comments which helped to 
improve the manuscript. We also thank Frank van den Bosch, 
Simon White, Ben Moore, Michael Balogh and Gabriella De Lucia 
for helpful comments
and suggestions, and Cheng Li, Marijn Franx, 
Joop Schaye, Stefania Giodini, Marcel Haas
and Alexander Hansson for useful discussion.

    T.L.\ and J.J.\ are supported within the framework of the Excellence
    Initiative by the German Research Foundation (DFG) through the Heidelberg
    Graduate School of Fundamental Physics (grant number GSC 129/1).
    H.T.M.\ is supported by the DFG through grant LI 1801/2-1.
J.J.\ acknowledges support by the Gottlieb Daimler and Karl Benz
Foundation.

     SQL databases containing the full galaxy data for the SAM of G11
    at all redshifts and for both the Millennium and Millennium-II 
    simulations are publicly released at 
    \texttt{http://www.mpa-garching.mpg.de/millennium}. The Millennium site was created as 
    part of the activities of the German
    Astrophysical Virtual Observatory. 
    
    Funding for the SDSS and SDSS-II has been provided by the Alfred
    P.\ Sloan Foundation, the Participating Institutions, the National
    Science Foundation, the U.S.\ Department of Energy, the National
    Aeronautics and Space Administration, the Japanese Monbukagakusho,
    the Max Planck Society, and the Higher Education Funding Council
    for England. The SDSS Web Site is \texttt{http://www.sdss.org/}. 
    The SDSS is managed by the Astrophysical Research Consortium for
    the Participating Institutions. The Participating Institutions are
    the American Museum of Natural History, Astrophysical Institute
    Potsdam, University of Basel, University of Cambridge, Case
    Western Reserve University, University of Chicago, Drexel
    University, Fermilab, the Institute for Advanced Study, the Japan
    Participation Group, Johns Hopkins University, the Joint Institute
    for Nuclear Astrophysics, the Kavli Institute for Particle
    Astrophysics and Cosmology, the Korean Scientist Group, the
    Chinese Academy of Sciences (LAMOST), Los Alamos National
    Laboratory, the Max-Planck-Institute for Astronomy (MPIA), the
    Max-Planck-Institute for Astrophysics (MPA), New Mexico State
    University, Ohio State University, University of Pittsburgh,
    University of Portsmouth, Princeton University, the United States
    Naval Observatory, and the University of Washington.

    This research has made use of the VizieR catalogue access
    tool, CDS, Strasbourg, France, of NASA's Astrophysics Data
    System Bibliographic Services, of the NASA/IPAC Extragalactic
    Database (NED) which is operated by the Jet Propulsion Laboratory, California
    Institute of Technology, under contract with the National Aeronautics
    and Space Administration, and of the ``K-corrections calculator'' service available at \texttt{http://kcor.sai.msu.ru/}.

\appendix

\section{Observational galaxy samples of nearby clusters}
\label{app:obs}

\subsection{The Virgo cluster}
Our initial Virgo cluster sample contains all galaxies of the Virgo Cluster
Catalog (VCC, Binggeli et al. 1986) that have $B$-magnitudes $m_B\le18.0$
(completeness limit of the VCC) and are either certain or possible
cluster members. This membership, initially based largely on morphology, was
updated by Binggeli et al. (1993) and by one of us
(T.L.) in May 2008 through velocities given by NED (Schneider et al.\ 1990, de
Vaucouleurs et al.\ 1991, Strauss et al.\ 1992, Lu et al.\ 1993,
Oosterloo et al.\ 1993, Zabludoff et al.\ 1993, Bettoni \& Galletta
1994, di Nella et al.\ 1995, Fisher et al.\ 1995, Kenney et al.\ 1995,
Rand 1995, Young \& Currie 1995, Drinkwater et al.\ 1996, Giovanelli
et al.\ 1997, 2007, Simien \& Prugniel 1997, 2002, Grogin et
al.\ 1998, Falco et al.\ 1999, Gavazzi et al.\ 1999, 2000, 2004, 2006,
Smith et al.\ 2000, Trager et al.\ 2000, van Driel et al.\ 2000,
Conselice et al.\ 2001, Bernardi et al.\ 2002, Caldwell et al.\ 2003,
Cortese et al.\ 2003, Geha et al.\ 2003, Makarov et al.\ 2003, Paturel
et al.\ 2003, Wegner et al.\ 2003, Denicolo et al.\ 2005,
Adelman-McCarthy et al.\ 2007, Chilingarian et al.\ 2007, Evstigneeva
et al.\ 2007).
Galaxy classification for early-type dwarfs was updated following
Lisker et al. (2007), and a number of uncertain, ``amorphous'' or ambiguous
objects were reclassified by one of us (T.L.) based on SDSS images.
Galaxies with $v_{\rm helio} \ge
3500\,$km/s were excluded; the remaining galaxies have velocities
of $-730 \le v_{\rm helio} \le 2990\,$km/s
\footnote{Galaxies classified as background galaxies by Binggeli et al. (1985, 1993) were not included in our sample even when their newly available velocities fall in the given range. This concerns three objects that would otherwise be included in our final working sample.}.
 With the systemic
velocity of the cluster at $\sim1200\,$km/s, this corresponds to a
velocity range of $\pm2000\,$km/s.
We include also the possible members of the VCC, since many of these belong to
substructure (e.g.\ ``clouds'', see Binggeli et al. 1987) located in
front or behind the main cluster, which would mostly also be included
in the SAM due to the fact that we simply apply a lower and upper
velocity limit there. 1030 galaxies now remain in our sample, of
  which 748 are spectroscopically confirmed members.

Total $r$-band magnitudes and colours from $ugriz$-bands were measured
by Lisker et al. (2007), Janz \& Lisker (2009), and Meyer et al.\ (in prep.) on
SDSS data release 5 images (Adelman-McCarthy et al. 2007), using a proper sky
subtraction method (described in Lisker et al. 2007) that avoids the serious
overestimation of the sky by the SDSS pipeline
for nearby galaxies of large apparent size.
 All values are corrected for Galactic
extinction (Schlegel et al. 1998).
We adopt elliptical apertures corresponding
to two Petrosian semimajor axes (Petrosian 1976) for our total magnitudes,
and correct early-type galaxies at bright and intermediate
luminosities for the flux that is missing due to this approach when their
S\'ersic index (S\'{e}rsic 1963) is larger than 1 (Graham et al. 2005;
Janz \& Lisker 2008).
Colours were measured within an elliptical aperture corresponding to two
half-light semimajor axes for the early-type galaxies, and to one Petrosian semimajor axis for the
late types. These apertures are roughly
similar for an exponential radial profile.

\begin{figure}
\includegraphics[width=84mm]{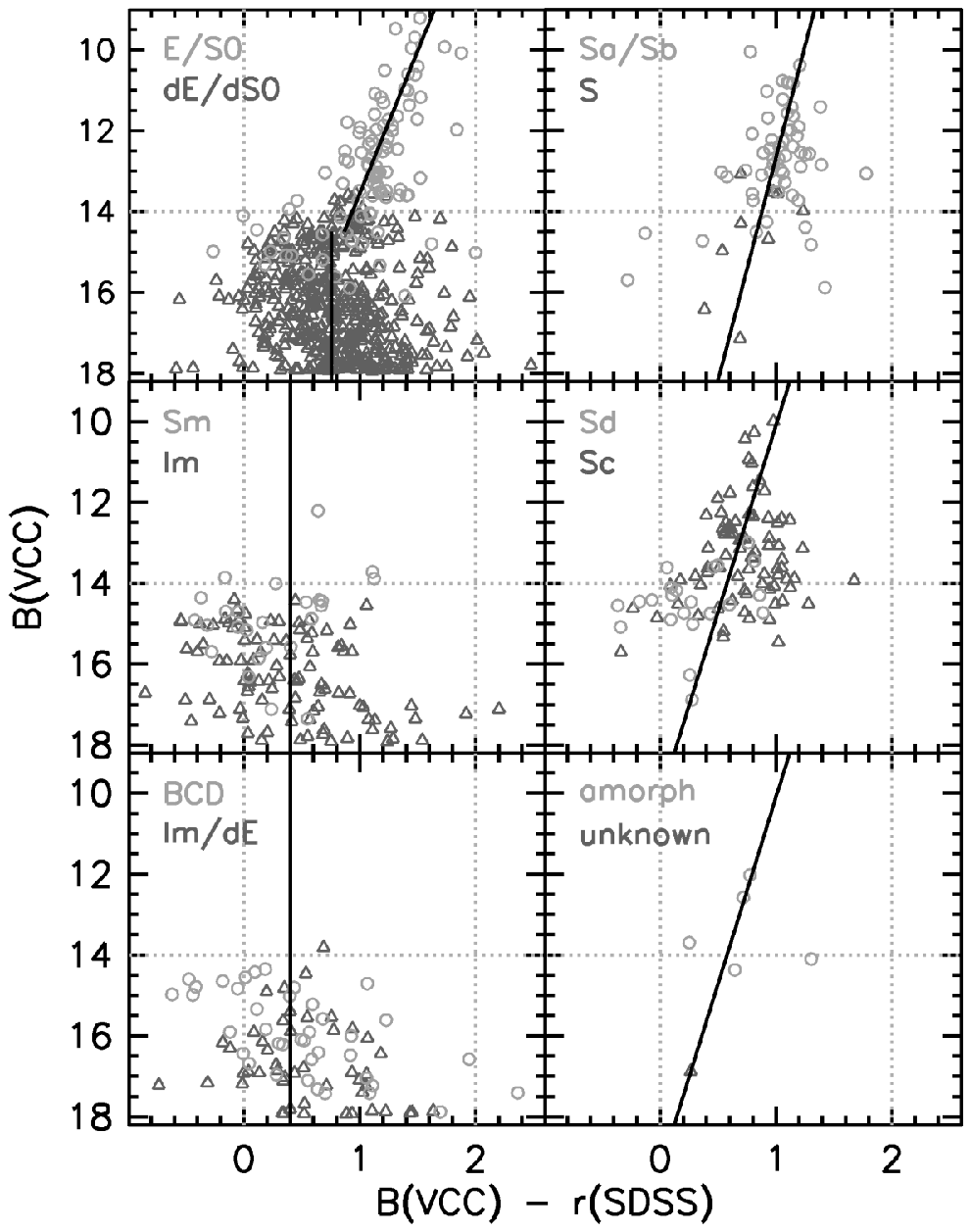}
 \caption[]{Transformations from VCC $B$-magnitudes to SDSS
   $r$-magnitudes, depending on galaxy type. ``S'' denotes spiral
   galaxies with unknown or ambiguous morphological subtype. Black
   lines show our adopted relations: $B-r = -0.1417 \cdot B + 2.9203$ for
   early types with $B\le14.5$, $B-r = 0.7570$ for early types with
   $B>14.5$;
   $B-r = -0.0907 \cdot B + 2.1489$ for classes Sa, Sb, and S;
   $B-r = -0.1082 \cdot B + 2.0925$ for classes Sc, Sd, ``amorphous'' and
   for the one unclassifiable galaxy;
   $B-r = 0.4010$ for classes Sm, Im, Im/dE, and BCD.}
 \label{fig:Br_trafo}
\end{figure}

For 84 galaxies, SDSS images were not or not successfully analyzed by
the named studies. For these, we obtained an estimate for their total
$r$-magnitude by applying a type-dependent $B-r$ transformation that
was found empirically from the 946 galaxies that were successfully
analyzed with SDSS data (Fig.~\ref{fig:Br_trafo}). 
The VCC $B$-magnitudes were corrected for Galactic extinction (Schlegel
et al. 1998)
before deriving the transformations.

We use a Virgo cluster distance modulus of $m-M=31.09$\,mag
(Mei et al. 2007; Blakeslee et al. 2009) for all galaxies, corresponding to $d=16.5$\,Mpc. With
the adopted WMAP1 cosmology, this leads to a spatial scale of
$0.079\,{\rm pc/''}$ or $0.286\,{\rm Mpc/^\circ}$. 
To translate the $B$-band completeness
  limit of $M_B\le -13.1$\,mag into an $r$-band completeness limit, we
  need to consider the following three aspects:
  (i) the errors in the $B$ and $r$-magnitudes, (ii) the differences of total
  $B$ and $r$ magnitudes, (iii) the empirical $B−to-r$ transformation
  adopted for 84 galaxies (see above). These issues are not
  independent of each other -- e.g.\ the errors in $B$ affect the
  distribution of $B-r$ values -- which we need to take into account
  as well. For aspect (i), we adopt a $B$-uncertainty of 0.4\,mag for
  the fainter VCC galaxies (Binggeli et al.\ 1985), and assume that
  the $r$-uncertainty is negligible compared to that (cf.\ Lisker et
  al.\ 2007). This uncertainty causes an additional, non-negligible
  scatter in the $B-r$ values (Fig.~A1). Considering this fact, we
  assume for aspect (ii) that the actual $B-r$ values reach up to
  1.5\,mag. Finally, while aspect (iii) introduces an additional
  uncertainty with respect to the sample completeness in $r$, the
  percentage of galaxies to which the empirical $B-r$ transformation
  was applied is small, and only a fraction of them would erroneously
  be moved above or below the completeness limit. We therefore assume
  that this has an effect of less than 0.2\,mag on the completeness
  limit.
  Taken together, this leads to an $r$-band completeness of
  $M_r\le-15.2$\,mag ($-13.1-0.4-1.5-0.2$).
We note
that with deep imaging of the Virgo cluster core region, Lieder et
al.\ (2011, in prep.) find no galaxies brighter than $M_r<-13$\,mag
that have not yet been identified by the VCC, confirming that our
completeness limit is reliable and probably rather
  conservative.
Due to the sky coverage of the VCC, our sample is spatially complete out
to a projected clustercentric distance (calculated from the central galaxy
M\,87) of 1.5 Mpc. Note that, due to this limitation, we omit the southern
subcluster around M\,49.
We adopt these limits for our final working sample, 
which contains 511 galaxies.

\subsection{The Fornax cluster}

The Fornax Cluster Catalog (FCC, Ferguson 1998), based on data with the
same instrument and similar quality as the VCC, provides us with an
initial Fornax cluster sample. It includes all 340 galaxies that are
certain and likely cluster members (Table II of Ferguson 1998), based
mainly on morphology, and lie within the
completeness limit of $m_B\le18.0$.

No SDSS data are available for the Fornax cluster. In order to obtain
estimated $r$-magnitudes, we apply the type-dependent $B-r$ transformation that
was found empirically for the Virgo cluster galaxies (Fig.~\ref{fig:Br_trafo}).
We rely on the fact that galaxy classification was performed
in a very similar manner in the VCC and FCC, and we assume that galaxy
colours are distributed similar in both clusters for a given galaxy type.
The FCC $B$-magnitudes were corrected for Galactic
extinction (Schlegel et al. 1998) before applying the transformations.

We use a Fornax cluster distance modulus of $m-M=31.51$\,mag
(Blakeslee et al. 2009) for all galaxies, corresponding to $d=20.0$\,Mpc. With
the adopted WMAP1 cosmology, this leads to a spatial scale of
$0.096\,{\rm kpc/''}$ or $0.346\,{\rm Mpc/^\circ}$.
To estimate an $r$-band completeness limit from the $B$-band
  completeness ($M_B\le -13.5$\,mag), the same three aspects apply as
  outlined in Sect.~A1 for 
  Virgo, with a slightly smaller $B$-uncertainty of 0.3\,mag (Ferguson
  1989). However, since for Fornax \emph{all} $r$-magnitudes are
  derived through the empirical $B-r$ transformations, aspect (iii) from
  above now becomes significant: the uncertainties in the adopted
  transformations can be as large as $\sim 0.6$\,mag for the bulk of
  data points (Fig.~A1).  This leads
  to an $r$-band completeness of $M_r\le-15.9$\,mag ($-13.5-0.3-1.5-0.6$).
Due to the sky coverage of the FCC, our sample is spatially complete out
to a projected clustercentric distance (calculated from the central galaxy
NGC\,1399) of 0.9 Mpc.
We adopt these limits for our final working sample, 
which contains 76 galaxies.

\subsection{The Coma cluster}

We use two different observational samples for the Coma cluster, which
we find to be in good agreement. The first sample (named ``Coma''
hereafter) is given by Michard \& Andreon (2008), based on spectroscopic and
morphological membership criteria and covering the central area of the
cluster. The second sample (``ComaB'') is constructed from SDSS data
and, in addition to spectroscopic member galaxies, involves a
statistical correction for the number of contaminating background
galaxies.

Michard \& Andreon (2008) provide a list of 473 Coma member galaxies,
initially based on the Godwin et al. (1983) catalog.
 We unambiguously
identified all but 7 galaxies in the SDSS data release 7
(Abazajian et al. 2009). Using SDSS spectroscopy, we apply
a redshift limit of $4000\le cz \le10\,000\,$km/s (cf.\
Kent \& Gunn 1982 and Chiboucas et al. 2010), within the given redshift
errors and using only redshifts with a confidence of 95\% or
higher. These limits remove 4 galaxies from the sample.

As total $r$-magnitudes, we adopt the SDSS photometric pipeline
Petrosian magnitudes, which are also used for $ugriz$-colours. We
checked that the above-mentioned overestimation of the sky
background that occurs for Virgo cluster galaxies does not have
significant effect for Coma, which simply is due to the much smaller
apparent size of the galaxies. Still, the luminosity of the two
massive central ellipticals NGC\,4874 and NGC\,4889 is clearly underestimated in the SDSS. For
them, as well as for the 7 galaxies not identified in the SDSS, we
obtain estimated $r$-magnitudes by applying the transformation of
Smith et al. (2002) to the Godwin et al. (1983) $B$ and $R$-magnitudes as
given by Michard \& Andreon (2008). 
All values are corrected for Galactic
extinction (Schlegel et al. 1998).

We use a Coma cluster distance modulus of $m-M=35.00$\,mag
(Carter et al. 2008) for all galaxies, corresponding to $d=100.0$\,Mpc. With
the adopted WMAP1 cosmology, this leads to a spatial scale of
$0.463\,{\rm kpc/''}$ or $1.667\,{\rm Mpc/^\circ}$.
With the Godwin et al.\ (1983) magnitude completeness of $m_B
  \lesssim 20.0$\,mag (Michard \& Andreon 2008), the $B$-uncertainty
  of 0.2\,mag (Godwin et al.\ 1983), and the $B-r$ values of up to
  1.5\,mag (see Sect.~A1), completeness in the r-band is only
  guaranteed down to $M_r = −16.7$\,mag.
We define the cluster center to be located midway between the two
central ellipticals, at $\alpha=194.9668^\circ,
\delta=+27.9680^\circ$. This leads to a spatial completeness out to a
projected clustercentric distance of 0.5 Mpc. These limits lead to a
final ``Coma'' working sample of 226 galaxies.

For the ``ComaB'' sample, based purely on the SDSS, we first select
all objects identified by the SDSS as galaxies within $2.5^\circ$
($4.2$\,Mpc) from the cluster center as defined above, with Petrosian
$r$-magnitudes (corrected for Galactic extinction) corresponding to
$M_r\le-16.7$\,mag at the Coma distance.
Where
SDSS spectra are available, with a redshift confidence of 95\% or
higher, we apply the above redshift limits. We then visually inspect
all objects and exclude them if they are, (i) stars, (ii) ``pieces'' of a galaxy
instead of stand-alone objects, or (iii) small
objects that are obviously too faint for our criteria, but whose
magnitude was artificially increased by the halo of a bright star or
the outer parts of a bright galaxy. A plot of magnitude versus surface
brightness helped to identify, as clear outliers, the few objects that fulfill one of the
above criteria but were overlooked during the first visual
inspection. Finally, we exclude all objects redder than $g-i>1.33$,
since these are even redder than all of the massive cluster
ellipticals, which should have the oldest and most metal-rich stellar populations among
the cluster members. These red, mostly faint galaxies almost certainly are
background contaminants at higher redshift. The final ``ComaB'' sample
contains 2141 galaxies, of which 923 are spectroscopic members.

\begin{figure}
\includegraphics[width=84mm]{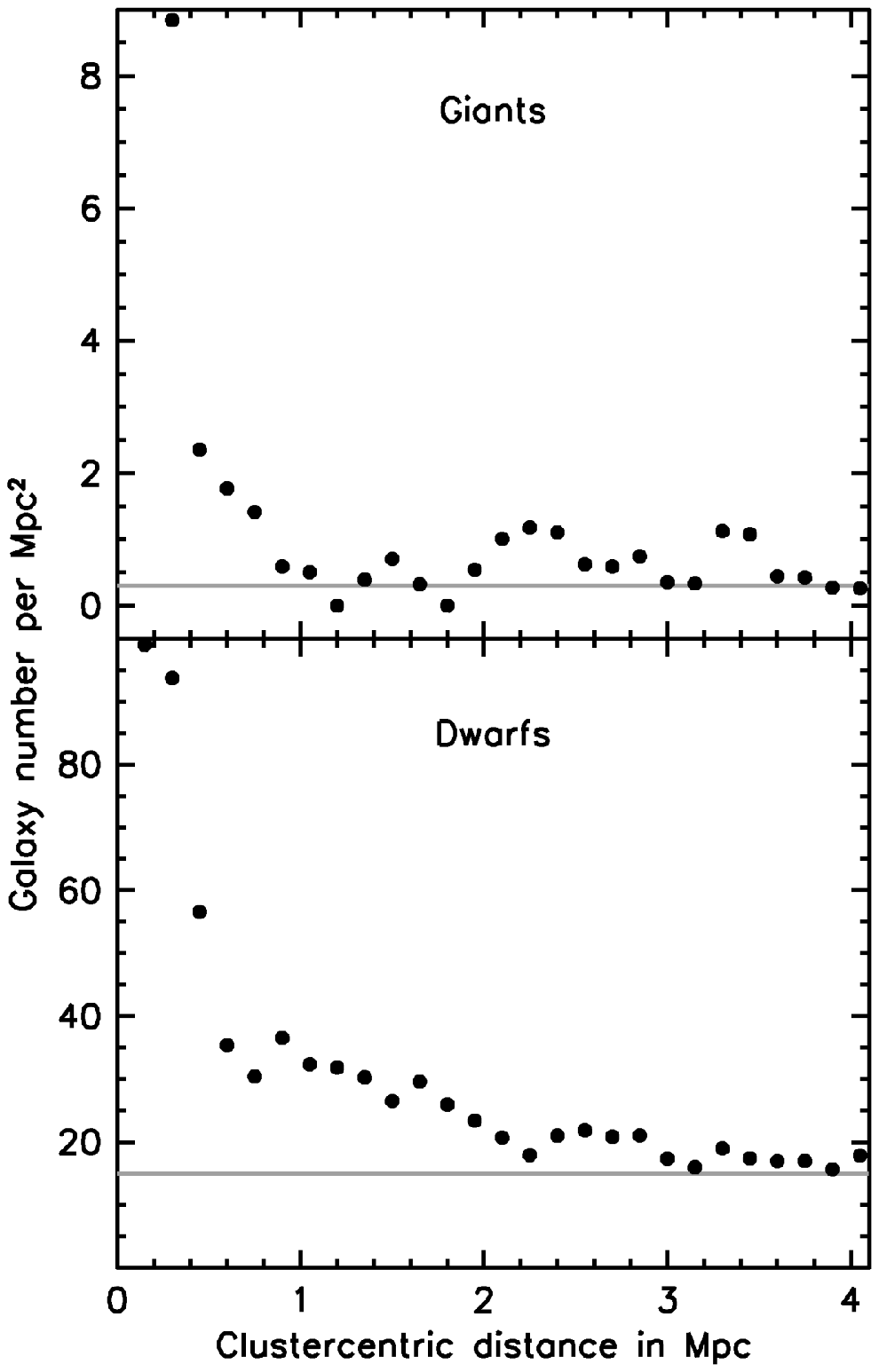}
 \caption[]{Radial number density of our ComaB sample, separated into
   dwarfs and giants at $M_r=-19$\,mag, and sampled with bins of
   0.3\,Mpc width in steps of 0.15\,Mpc. Only galaxies without
   spectroscopic membership information are included. The grey
   horizontal lines denote our adopted background values.}
 \label{fig:comabg}
\end{figure}

From a radial number density profile of those
galaxies that are not spectroscopic members, we decide to adopt a statistical
background galaxy number density of $15\,{\rm Mpc^{-2}}$ for dwarfs
($-19.0\le M_r \le -16.7\,$mag), and  $0.3\,{\rm Mpc^{-2}}$ for giants
($M_r<-19.0\,$mag; see Fig.~\ref{fig:comabg}). Therefore, in a statistical sense, 835 galaxies of
the sample are background objects, and 383 are cluster members without
an SDSS redshift.
We remark 
that the SDSS spectroscopic coverage reaches only to
$M_r\lesssim-17.3$\,mag.

To estimate the red fraction, we first k-correct the $g-r$ colours
of all galaxies to z=0, assuming they are all at a redshift of z=0.023, which corresponds
to the estimated distance to Coma. This is done using the approximation
of Chilingarian et al. (2010). We then
 background-correct red galaxies, defined
according to eq. \ref{eq:colour}, and galaxies in different 
absolute magnitude bins, separately.
We use zero background correction for galaxies with $M_r<-20$, 
and 
$0.5\,{\rm Mpc^{-2}}$, 
$2\,{\rm Mpc^{-2}}$ 
and $14\,{\rm Mpc^{-2}}$
for galaxies with 
$-20 \leq M_r \leq -18.5$,  
$-18.5 \leq M_r \leq -17.3$
and  $-17.3 \leq M_r \leq -16.7$ 
respectively. For red galaxies, 
we use a background correction of  $11\,{\rm Mpc^{-2}}$ for 
dwarfs, and of zero, $0.3\,{\rm Mpc^{-2}}$, $0.7\,{\rm Mpc^{-2}}$ and $
8.5\,{\rm Mpc^{-2}}$ for the four magnitude bins listed above.

For comparison, we also constructed an SDSS sample analogous to ComaB, but
only for the area covered by (Michard \& Andreon 2008). When assuming that
Michard \& Andreon 2008 correctly identified all cluster members, we would
obtain a background contamination value of $23\,{\rm Mpc^{-2}}$ for
dwarfs, i.e.\ a factor $\sim1.5$ larger than our adopted value for the
ComaB sample. However, there are two indications that this value might
indeed be too large (aside from the general possibility of
strong cosmic variance). First, if we would use $23\,{\rm Mpc^{-2}}$
for the ComaB sample, then \emph{all} dwarfs without SDSS spectra
and even $\sim$70 of those with spectroscopically confirmed
membership would statistically have to be background
contaminants. Second, Chiboucas et al. 2010 present spectroscopic
membership for a number of faint Coma dwarfs also included in
Michard \& Andreon (2008), with magnitudes even fainter than their
completeness limit, and find that Michard \& Andreon (2008) erroneously
assigned a substantial fraction of spectroscopic cluster members to
the background population, based on their morphological
criteria. While these criteria probably work better at the somewhat
brighter magnitudes
that we are concerned about, this seems at least consistent with our findings.
It should be noted that, as far as the \emph{central} cluster region
is concerned, a small variation in the background correction
does not have a significant effect on the number density profile 
(see Fig.\ref{fig:v1}), since the vast majority of galaxies in that region are
  cluster members anyway.

\subsection{The Perseus cluster}

 We construct a Perseus cluster sample from SDSS data release 7
(Abazajian et al. 2009) similar to the ComaB sample, except for the fact
that there are no SDSS spectra available for this region, and that the
spatial coverage in the cluster outskirts is incomplete. We first select
all objects identified by the SDSS as galaxies within $3.0^\circ$
($3.8$\,Mpc) from the cluster center, taken to be the
central massive galaxy NGC\,1275.
As for Coma, we adopt a magnitude limit of $M_r\le-16.7$\,mag at the Perseus distance
(see below), already corrected for Galactic extinction and using SDSS
Petrosian $r$-magnitudes. Since Perseus is closer than Coma, adopting the same limit in
  \emph{absolute} magnitude means a shallower limit in \emph{apparent}
  magnitude, which is useful to keep the amount of background
  contaminants at a moderate level.

The spatial incompleteness of the SDSS can
be characterized as follows. Coverage is complete
out to $0.75^\circ$ ($0.95$\,Mpc) from the cluster center. For the sample galaxies beyond $0.75^\circ$ and out to
$2.0^\circ$ ($2.5$\,Mpc), the average completeness correction factor is
1.50, with a standard deviation of 0.23. The SDSS thus still covers
more than half of the cluster area in the outskirts, and should thus
be well representative for the whole cluster. An additional small area
has been excluded (around $\alpha=50.17^\circ, \delta=+43.08^\circ$), since
a confirmed background cluster is located there.

As for the ComaB sample, we then visually inspect the Perseus sample
and exclude, (i) stars, (ii) ``pieces'' of a galaxy
instead of stand-alone objects, and (iii) small
objects that are obviously too faint for our criteria, but whose
magnitude was artificially increased by the halo of a bright star or
the outer parts of a bright galaxy.
However, in contrast to the ComaB selection process, we do not apply a
redward colour cut here, since the Perseus cluster is located at low
Galactic latitude ($b=-13.26^\circ$) and thus there might be local
extinction variations not taken into account in our adopted
Schlegel et al. (1998) values, affecting galaxy colours.
We are therefore left with 1853 galaxies in our working sample.

We use a Perseus cluster distance modulus of $m-M=34.29$\,mag,
corresponding to a "Hubble flow distance" of $d=72.3$\,Mpc. This is
given by NED based on the heliocentric velocity of $5366\,$km/s
(Struble \& Rood 1999), on WMAP1 cosmology, and on considering the
influence of the Virgo cluster, the Great Attractor, and the Shapley
Supercluster through the local velocity field model of
Mould et al. (2000).
This leads to a spatial scale of
$0.350\,{\rm kpc/''}$ or $1.260\,{\rm Mpc/^\circ}$, similar to what
previous studies used (Conselice et al. 2002; Sanders \& Fabian 2007).

\begin{figure}
 \includegraphics[width=84mm]{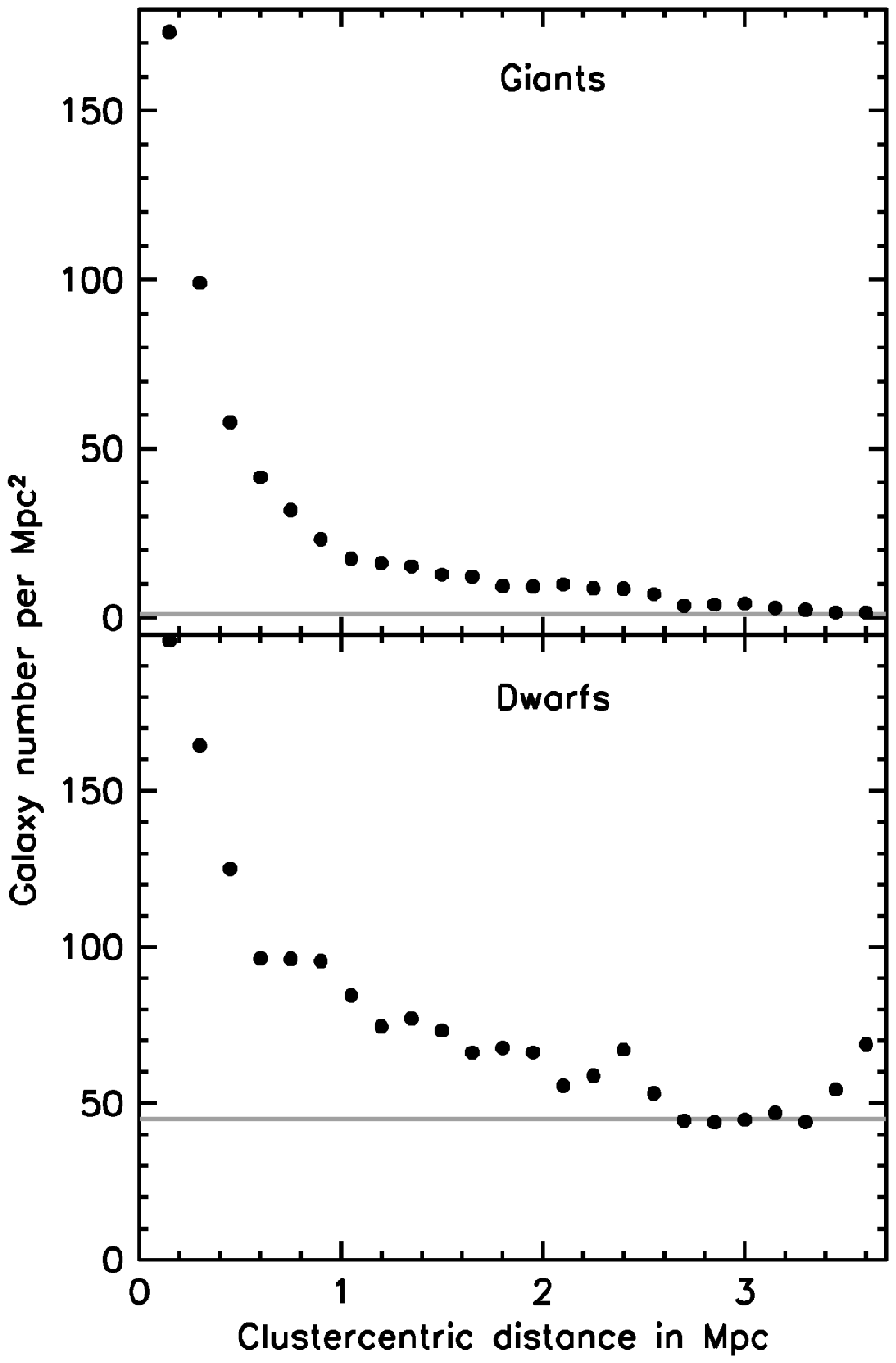}
 \caption[]{Radial number density of our Perseus sample, separated into
   dwarfs and giants at $M_r=-19$\,mag, and sampled with bins of
   0.3\,Mpc width in steps of 0.15\,Mpc. The grey
   horizontal lines denote our adopted background values.}
 \label{fig:perseusbg}
\end{figure}

From a radial number density profile, we find a statistical
background galaxy number density of $45\,{\rm Mpc^{-2}}$ for dwarfs
($-19.0\le M_r \le -16.7\,$mag) and  $1.0\,{\rm Mpc^{-2}}$ for giants
($M_r<-19.0\,$mag; see Fig.~\ref{fig:perseusbg}). While these values cannot be determined to high
accuracy, their uncertainty has little effect on the number densities
in the inner cluster regions.
To estimate the red fraction, we
 also background-correct red galaxies, defined
according to eq. \ref{eq:colour}, and galaxies in different 
absolute magnitude bins, separately. We k-correct Perseus galaxies
to z=0
using the approximation of Chilingarian et al. (2010), assuming they are all 
at a fixed redshift z=0.017.
We use zero background correction for galaxies with $M_r<-20$, 
and 
$4\,{\rm Mpc^{-2}}$, 
$18\,{\rm Mpc^{-2}}$ 
and $25\,{\rm Mpc^{-2}}$
for galaxies with 
$-20 \leq M_r \leq -18.5$,  
$-18.5 \leq M_r \leq -17.3$
and  $-17.3 \leq M_r \leq -16.7$ 
respectively. For red galaxies, 
we use a background correction of  $37\,{\rm Mpc^{-2}}$ for 
dwarfs, and of zero, $3\,{\rm Mpc^{-2}}$, $14\,{\rm Mpc^{-2}}$ and $
21\,{\rm Mpc^{-2}}$ for the four magnitude bins listed above.

\subsection{Completeness for compact objects}

While published magnitude completeness limits (e.g.\ VCC, FCC) primarily relate
to losing galaxies with very low surface brightness on the diffuse
side, the completeness on the compact side needs to be considered as
well. Very compact galaxies might be confused with stars when they
are barely resolved, or be confused with background galaxies due to
their small radius and comparably high surface brightness. The
compilation of Misgeld et al.\ (2011, their Fig.~1) shows that the
smallest known galaxies down to our completeness limits can have
effective radii as small as $\sim$100\,pc. Note that the
so-called ultra-compact dwarf (UCD) galaxies are fainter than our limits.

We can reasonably assume that such objects can be seen on their
respective imaging data (VCC, FCC, SDSS, Michard \& Andreon 2008) at
least out to two effective radii, i.e.\ have a visible diameter of at
least $\sim$400\,pc. This corresponds to $\sim$5 arcseconds at the Virgo
cluster distance, and $\sim$4 arcseconds at the Fornax cluster
distance. Clearly, all such galaxies would be recognized as extended
objects in Virgo and Fornax. For the larger distances of the Coma and
Perseus clusters, the situation changes: the named galaxies would have
visible diameters of only 0.9 and 1.1 arcseconds, respectively, or
between 2 and 3 SDSS pixels. However, the fraction of galaxies with
effective radii below 400\,pc (i.e.\ a visible diameter below
1.6\,kpc) is at most a few percent (Janz \& Lisker 2008), too small to
affect any of our conclusions.

\end{document}

%% file: macros_wl.tex
\def\beq{\begin{equation}}
\def\eeq{\end{equation}}
\def\barray{\begin{eqnarray}}
\def\earray{\end{eqnarray}}

\def\gtsima{$\; \buildrel > \over \sim \;$}
\def\ltsima{$\; \buildrel < \over \sim \;$}
\def\prosima{$\; \buildrel \propto \over \sim \;$}
\def\gsim{\lower.7ex\hbox{\gtsima}}
\def\lsim{\lower.7ex\hbox{\ltsima}}
\def\simgt{\lower.7ex\hbox{\gtsima}}
\def\simlt{\lower.7ex\hbox{\ltsima}}
\def\simpr{\lower.7ex\hbox{\prosima}}
\def\la{\lsim}
\def\ga{\gsim}

\newcommand{\apj}{ApJ}
\newcommand{\apjs}{ApJS}
\newcommand{\apjl}{ApJL}
\newcommand{\aj}{AJ}
\newcommand{\mnras}{MNRAS}
\newcommand{\aap}{A\&A}
\newcommand{\aaps}{A\&AS}

\newcommand{\nat}{Nature}

\newdimen\hssize
\newdimen\hdsize
\def\fsdb07{f_{\rm s,D}}